\documentclass[journal]{IEEEtran}   

\usepackage{amsmath,amssymb,mathtools}
\usepackage{bm}        
\usepackage{amsthm}    
\usepackage{svg}
\usepackage{booktabs}
\usepackage{titlesec}
\DeclareMathOperator{\diag}{diag}

\newcommand{\CC}{\mathbb C}

\newcommand{\CN}{\mathcal{CN}}

\theoremstyle{plain}

\theoremstyle{definition}

\theoremstyle{remark}

\usepackage{xcolor}

\usepackage{graphicx}
\usepackage[caption=false,font=footnotesize]{subfig}

\usepackage[section]{placeins}

\usepackage{algorithm}
\usepackage{algpseudocode}

\algrenewcommand\algorithmicrequire{\textbf{Input:}}
\algrenewcommand\algorithmicensure{\textbf{Output:}}

\usepackage{comment}
\usepackage{url}
\usepackage{cite}

\usepackage[acronym,shortcuts]{glossaries}
\makeglossaries

\usepackage[hidelinks]{hyperref}

\usepackage{amsmath,amssymb,mathtools}

\usepackage{bm}

\newcommand{\Normal}{\mathcal{N}}   

\newacronym{LDPC}{LDPC}{low-density parity check}
\newacronym{ISFFT}{ISFFT}{inverse symplectic finite Fourier transform}
\newacronym{SFFT}{SFFT}{symplectic finite Fourier transform}
\newacronym{DFT}{DFT}{discrete Fourier transform}
\newacronym{DoF}{DoF}{degrees of freedom}
\newacronym{2D}{2D}{two-dimensional}
\newacronym{1D}{1D}{one-dimensional}
\newacronym{3GPP}{3GPP}{3rd generation partnership project}
\newacronym{5G}{5G}{fifth generation}
\newacronym{5GNR}{5GNR}{fifth generation new radio}
\newacronym{6G}{6G}{sixth-generation}
\newacronym{ADC}{ADC}{analog-to-digital converter}
\newacronym{AMP}{AMP}{approximate message passing}
\newacronym{AWGN}{AWGN}{additive white Gaussian noise}
\newacronym{AoA}{AoA}{angle of arrival}
\newacronym{AoD}{AoD}{angle of departure}
\newacronym{BP}{BP}{belief propagation}
\newacronym{BS}{BS}{base station}
\newacronym{CRLB}{CRLB}{Cram\'{e}r--Rao lower bound}
\newacronym{CSI}{CSI}{channel state information}
\newacronym{GaBP}{GaBP}{Gaussian belief propagation}
\newacronym{EM}{EM}{electromagnetic}
\newacronym{FDTD}{FDTD}{finite-difference time-domain}
\newacronym{ISAC}{ISAC}{integrated sensing and communication}
\newacronym{LSM}{LSM}{linear sampling method}
\newacronym{LiDAR}{LiDAR}{light detection and ranging}
\newacronym{MAP}{MAP}{maximum \textit{a posteriori}}
\newacronym{MIMO}{MIMO}{multiple-input multiple-output}
\newacronym{NMSE}{NMSE}{normalized mean square error}
\newacronym{NG-RAN}{NG-RAN}{next-generation radio access network}
\newacronym{QP}{QP}{quadratic programming}
\newacronym{ROI}{ROI}{region of interest}
\newacronym{SNR}{SNR}{signal-to-noise ratio}
\newacronym{UCA}{UCA}{uniform circular array}
\newacronym{ULA}{ULA}{ uniform linear array}
\newacronym{WSS}{WSS}{wide-sense stationary}
\newacronym{FIM}{FIM}{Fisher information matrix}
\newacronym{TDD}{TDD}{time division duplex}
\newacronym{BIM}{BIM}{Born iteration method}
\newacronym{DT}{DT}{digital twin}
\newacronym{PDP}{PDP}{power delay profile}
\newacronym{BF}{BF}{beam-forming}
\newacronym{OFDM}{OFDM}{orthogonal frequency-division multiplexing}
\newacronym{F2V}{F2V}{factor-to-variable}
\newacronym{V2F}{V2F}{variable-to-factor}
\newacronym{DTN}{DTN}{digital twin network}
\newacronym{CDF}{CDF}{cumulative distribution function}
\newacronym{TX}{TX}{transmit}
\newacronym{RX}{RX}{receive}

\begin{document}

\title{Electromagnetic Digital Twin-Enabled Closed-Loop Beam Management in ISAC Systems }

\author{%
Yubin Luo, Takumi Takahashi, \IEEEmembership{Member, IEEE}, Li Yu, \IEEEmembership{Member, IEEE}, Zhaohui Yang, \IEEEmembership{Member, IEEE},\\ Jianhua Zhang, \IEEEmembership{Fellow, IEEE} and Hideki Ochiai, \IEEEmembership{Fellow, IEEE}
    \thanks{Y. Luo, T. Takahashi and H. Ochiai are with the Graduate School of Engineering, The University of Osaka, 2-1-Yamada-oka, Suita, 565-0871, Japan. (e-mail: y-luo@wcs.comm.eng.osaka-u.ac.jp, takahashi@comm.eng.osaka-u.ac.jp, ochiai@comm.eng.osaka-u.ac.jp.)}
    \thanks{L. Yu, and J. Zhang are with the State Key Laboratory of Networking and Switching Technology, Beijing University of Posts and Telecommunications, Beijing 100876, China. (e-mail: li.yu@bupt.edu.cn; jhzhang@bupt.edu.cn.)}
    \thanks{Z. Yang is with the Zhejiang Key Laboratory of Information Processing Communication and Networking, College of Information Science and Electronic Engineering, Zhejiang University. (e-mail: yang\_zhaohui@zju.edu.cn.)}

\vspace{-5mm}
}

\maketitle

\begin{abstract}
\Ac{DT} is envisioned as a key enabler of \ac{6G} communication systems, evolving from offline descriptive replicas for monitoring and analysis to in-the-loop agents within \acp{DTN} that couple physical and digital worlds.
Recent advances in \ac{ISAC}-driven \ac{EM} scattering methods enable environment twinning by linking channel behaviors to \ac{EM} properties of the scatterers, supporting interpretable \ac{DT} states and \ac{EM}-grounded optimization.
However, existing studies primarily focus on \ac{DT} construction and lack mechanisms for closed-loop control in wireless systems.
Moreover, array-geometry mismatch can bias \ac{DT} reconstruction and degrade control performance, while prior works assume known arrays.
To address these gaps, we propose an \ac{EM}-\ac{ISAC}-based closed-loop \ac{DTN} framework with a hierarchical design integrating \textit{environment twinning, prior injection, and control decision} into an end-to-end loop.
Leveraging \ac{ISAC} measurements, the proposed framework jointly reconstructs scatterer information and array-dependent forward operator and employs a low-complexity Bayesian message-passing algorithm to perform contrast inference and array calibration.
The reconstructed \ac{DT} guides codebook preselection to reduce training overhead and narrow candidate beams.
Subsequently, downlink \ac{BF} is performed based on \ac{DT}-predicted channels, enabling latency-bounded closed-loop control.
Simulation results demonstrate improved robustness and control performance under array mismatch.

\end{abstract}
\glsresetall

\begin{IEEEkeywords}
Digital twin network, integrated sensing and communication, belief propagation, beam management.
\end{IEEEkeywords}
\ifCLASSOPTIONpeerreview
  \begin{center}\bfseries EDICS Category: 3-BBND \end{center}
\fi
\IEEEpeerreviewmaketitle

\section{Introduction}

\Ac{DT}, as a key vision for future \ac{6G} communication systems, is driving the evolution toward intelligent design and optimization in wireless networks~\cite{b1,b13}.
Traditionally, \ac{DT} provides a digital replica of the wireless environment and network dynamics, enabling interpretable channel-aware representations and informative priors for system optimization~\cite{b28}.
More recently, as wireless networks become reconfigurable and adaptive, \ac{DT} evolves from an offline, monitoring tool into an in-the-loop entity that bridges the physical and digital worlds, enabling closed-loop decision-making.
This transition leads to the concept of \acp{DTN}, which aims to achieve physically--digitally consistent and scalable system optimization~\cite{b2,b12}. 

Against this backdrop, existing \ac{DT}-enabled wireless solutions can be broadly categorized as strategies that couple \ac{DT} construction with closed-loop control, depending on what is twinned and how the twin is updated and exploited.
Specifically, \cite{b3} constructs a network \ac{DT} for reliable edge caching, where the physical network is mirrored in the digital domain via vertical and horizontal twinning to support reliability-aware decision-making.
In \cite{b4}, a machine-learning-based \ac{DT} is developed for predictive modeling in wind turbines and integrated with a \ac{NG-RAN}-assisted cloud platform, enabling virtual monitoring and closed-loop feedback.
In \cite{b5}, a large-model-based channel \ac{DT} is proposed for real-world environment intelligence, where multimodal data and learned representations are used to predict channel behaviors and facilitate downstream wireless optimization.
In \cite{b6}, \acp{DT} of \ac{EM} propagation environments are constructed for live \ac{5G} networks by passively acquiring multipath parameters from in-network signals and combining \ac{EM} simulation to reproduce site-specific propagation effects in the electrical domain.
Furthermore, in \cite{b7}, an \ac{EM}-scattering-based material reconstruction approach is proposed to reconstruct environmental scatterers using endogenous data from an \ac{ISAC} system.

Among these paradigms, the \ac{ISAC}-driven \ac{EM}-scattering perspective has recently attracted increasing attention for \ac{DT} construction, as it provides a physics-consistent view of wireless propagation rooted in \ac{EM} field interactions~\cite{b11,b12,b29}.
This paradigm is particularly appealing because it enables the construction of a physical \ac{DT} directly from endogenous data available in communication systems, without requiring additional sensing infrastructures.
In contrast, conventional approaches to physical-world twinning often rely on extra on-site sensors, such as \ac{LiDAR} and cameras, increasing deployment costs and system complexity.
By linking channel behaviors to the geometry and material properties of environmental scatterers, this approach enables reconstruction of a physically interpretable environment \ac{DT} state and supports \ac{EM}-grounded optimization.
For wireless systems, such a physics-consistent paradigm turns environment awareness into actionable priors for beam management and resource adaptation, bridging propagation modeling with communication and sensing design~\cite{b20,b28}.
However, most existing studies focus primarily on \ac{DT} construction, including pilot design~\cite{b8}, channel reconstruction~\cite{b9}, and analysis of reconstruction ill-conditioning~\cite{b10}.

A key challenge in closing the loop in practice is how to translate a reconstructed, potentially imperfect \ac{DT} state into reliable and latency-constrained wireless control actions. Achieving this goal requires trustworthy sensing and \ac{BF} models that remain robust under model mismatch and hardware imperfections~\cite{b14,b15,b16}. Furthermore, most existing \ac{DT}-enabled propagation studies assume perfectly known array geometry, under which the propagation channel and array geometry are fully determined once the environment \ac{DT} is obtained. In realistic settings, however, the array geometry is often unknown or mismatched, which can distort both the scattering operator and the array manifold, thereby biasing the reconstructed \ac{DT} and degrading the reliability of the resulting closed-loop control actions. Consequently, array calibration becomes a critical component of the overall pipeline.

\begin{figure}[!t]
  \centering
  \hspace*{-0.05\columnwidth}
  \includegraphics[width=1.07\columnwidth]{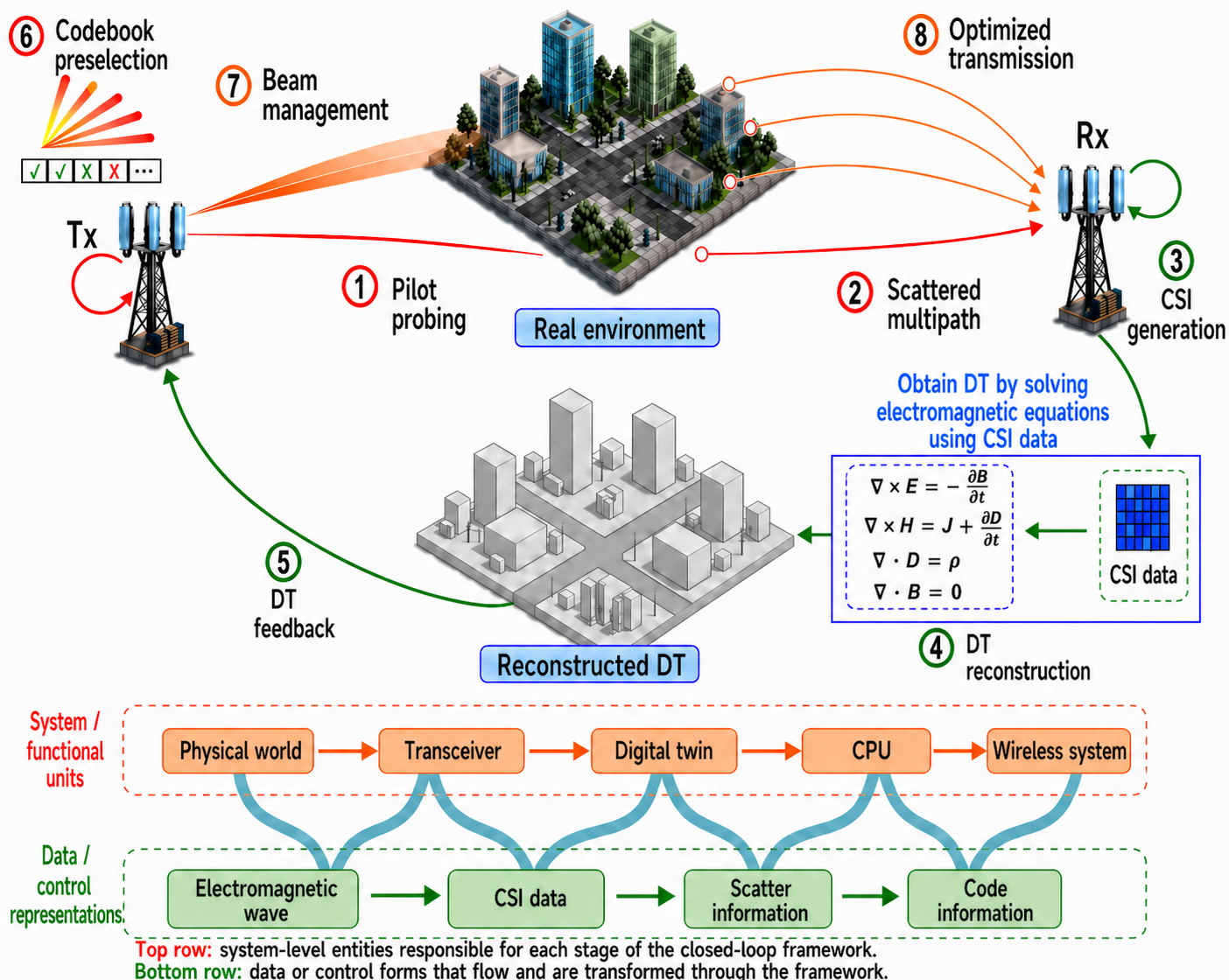}
  \caption{Overall workflow of EM-DT-enabled closed-loop
beam management.}
  \label{fig:system_model}
  \vspace{3.6mm}
\end{figure}

Motivated by the above gaps, we develop an electromagnetic digital-twin-based closed-loop \ac{DTN} framework realized through \ac{ISAC} systems.
Specifically, we move beyond environment twinning and focus on making the reconstructed \ac{DT} \emph{actionable} for latency-bounded wireless control under non-ideal hardware conditions.
To this end, we explicitly account for array calibration by jointly inferring the environment state and the array-dependent forward operator, enabling reliable downstream beam management and \ac{BF} decisions.

To operationalize this vision, we develop a unified inference-and-control framework that closes the loop from environment twinning to executable beam management.
The proposed framework consists of two tightly coupled stages: \ac{DT} reconstruction and closed-loop control.
In the first stage, \ac{DT} reconstruction aims to recover the physical information of the environment, such as the material properties of the scatterers and the geometry of the antenna array.
To address the large problem size, we adopt \ac{GaBP}~\cite{b17,b18,b18a} as a low-complexity Bayesian message-passing engine, with an alternating procedure that performs contrast inference and array calibration within a unified computational backbone, yielding an uncertainty-aware environment \ac{DT}.
In the second stage, closed-loop beam management is built upon the reconstructed \ac{DT}: \ac{DT}-guided codebook preselection reduces training overhead, while uncertainty-aware downlink \ac{BF} based on the \ac{DT}-predicted channel enables latency-bounded control with feedback. From a system-level perspective, the proposed closed-loop \ac{DTN} framework trades offline \ac{DT} construction and digital-domain inference for reduced over-the-air training and beam-search overhead.

The main contributions of this work are summarized as follows:
\begin{itemize}
\item We propose a unified inference framework that jointly performs environment reconstruction and array calibration via a \ac{GaBP}-driven pipeline, improving  reliability for physical \ac{DT} reconstruction under model mismatch.
\item We develop a \ac{DT}-based closed-loop beam management scheme that integrates \ac{DT}-guided codebook preselection with uncertainty-aware \ac{BF}.
\item Through numerical simulations, we demonstrate that the proposed closed-loop \ac{DTN} architecture can be realized entirely from endogenous communication-system data, enabling physical \ac{DT} reconstruction, array calibration, and downstream beam optimization within a unified in-the-loop framework.
\end{itemize}

The remainder of this paper is organized as follows.
Section~II introduces the \ac{EM}-\ac{ISAC} system model.
Section~III presents the proposed \ac{GaBP}-based \ac{DT} reconstruction framework for scatter reconstruction and array calibration.
Section~IV describes the closed-loop beam management design, including \ac{DT}-guided codebook preselection and uncertainty-aware \ac{BF} optimization.
Section~V presents numerical results, and Section~VI concludes the paper.

\textit{Notation:}
Boldface letters denote vectors and matrices.
$\mathrm{j}$ is the imaginary unit.
$(\cdot)^{\mathrm T}$, $(\cdot)^{\mathrm H}$, and $(\cdot)^*$ denote transpose, Hermitian transpose, and complex conjugate, respectively.
$\Re\{\cdot\}$, $\Im\{\cdot\}$, $|\cdot|$, $\|\cdot\|_2$, and $\langle\cdot,\cdot\rangle$ denote the real part, imaginary part, modulus, Euclidean norm, and inner product, respectively.
$\mathrm{diag}(\mathbf x)$ forms a diagonal matrix from $\mathbf x$; $\mathbf I_M$ is the $M\times M$ identity matrix; $\circ$ denotes the Khatri--Rao product; and $\mathbf A_{m,:}$ and $\mathbf A_{:,i}$ denote the $m$-th row and $i$-th column of $\mathbf A$.
$\mathbf x\sim\mathcal{CN}(\boldsymbol\mu,\mathbf V)$ and $x\sim\mathcal{CN}(\mu,v)$ denote complex Gaussian random vectors and variables, respectively, while $\mathcal N$ denotes the real Gaussian distribution.
$\mathbb E[\cdot]$, $\mathrm{Var}[\cdot]$, and $\Pr\{\cdot\}$ denote expectation, variance, and probability.
$\operatorname{eig}_{\max}(\mathbf A,\mathbf B)$ denotes the generalized eigenvector associated with the largest generalized eigenvalue of $(\mathbf A,\mathbf B)$. $\dim(\mathbf x)$ denotes the dimension of $\mathbf x$, and $\mathbf A\succeq0$ denotes $\mathbf A$ is positive semidefinite.

\section{System Model}

We consider a \ac{2D} transverse magnetic ($\mathrm{TM}_z$) \ac{EM} sensing scenario with an uplink \ac{MIMO}-\ac{OFDM} probing protocol.
The $\mathrm{TM}_z$ assumption is adopted for tractability, under which the \ac{EM} fields are invariant along the $z$-axis.
Let $\mathcal D\subset\mathbb R^2$ denote the sensing region, which is discretized into $M$ grid points, and let $\Gamma$ denote a closed observation curve in the far field of $\mathcal{D}$.
A co-located aperture with $L$ elements is deployed on $\Gamma$ and operates in \ac{TDD} mode, thus the same aperture alternates between transmission and reception.
In each \ac{TDD} sensing snapshot, all $L$ elements are simultaneously excited using a fully digital pilot vector, and an $L$-dimensional baseband observation is collected during the receive phase.
We assume a fully digital baseband architecture throughout.

We consider $K$ equally spaced subcarriers centered at $f_c$ with spacing $\Delta f$, i.e.,
\begin{equation}
f_k \triangleq f_c+\Bigl(k-\frac{K+1}{2}\Bigr)\Delta f,\qquad k=1,\ldots,K.
\label{eq:subcarrier_def}
\end{equation}
At subcarrier $k$, the received baseband signal is expressed as
\begin{equation}
\mathbf y_k=\mathbf H_k\mathbf x_k+\mathbf n_k,\qquad
\mathbf n_k\sim \mathcal{CN}(\mathbf 0,\sigma^2\mathbf I_{L}),
\label{eq:mimo_ofdm_k}
\end{equation}
where $\mathbf y_k\in\mathbb C^{L\times 1}$ denotes the received signal, 
$\mathbf x_k\in\mathbb C^{L\times 1}$ is pilot vector, 
$\mathbf H_k\in\mathbb C^{L\times L}$ represents the end-to-end scattering channel, 
$\mathbf n_k$ denotes the additive noise, $\mathbf 0\in\mathbb C^{L\times 1}$ is the all-zero vector, 
and $\sigma^2$ denotes the noise variance.
By reciprocity under \ac{TDD} operation, the same array is used for both transmission and reception in \ac{DT} construction stage.

\subsection{EM-Consistent Scattering Channel and Forward Operator}

We adopt an \ac{EM}-consistent scattering model based on the Lippmann–Schwinger equation to characterize the propagation channel, which relates the total field to the incident field and the material contrast within the sensing region.
Let $\omega_k=2\pi f_k$.
Denote by $E_k^i(\mathbf r)$ and $E_k^t(\mathbf r)$ the incident and total $\mathrm{TM}_z$ fields at pixel $\mathbf r\in \mathcal D$, respectively.
The Lippmann–Schwinger equation is given by~\cite{b26}:
\begin{equation}
E_k^{t}(\mathbf r)
= E_k^{i}(\mathbf r)
+\int_{\mathcal D} G_k(\mathbf r,\mathbf r')\, \chi(\mathbf r';\omega_k)\,E_k^{t}(\mathbf r')\,\mathrm d\mathbf r'.
\label{eq:ls_integral}
\end{equation}
Here, $\mathbf r'$ is an arbitrary pixel in $\mathcal D$ and $\chi(\mathbf r;\omega_k)$ denotes the complex material contrast, capturing both permittivity and conductivity variations:

\begin{equation}
\chi(\mathbf r;\omega_k)\triangleq
\varepsilon_r(\mathbf r)-1+\mathrm{j}\frac{\sigma_r(\mathbf r)}{\varepsilon_0\omega_k}.
\label{eq:contrast}
\end{equation}
where $\varepsilon_r(\mathbf r)$ denotes the relative permittivity, 
$\sigma_r(\mathbf r)$ denotes the electrical conductivity, and $\varepsilon_0$ is the vacuum permittivity. We assume $\chi(\mathbf r;\omega_k)\approx \chi(\mathbf r;\omega_c)$ with $\omega_c \triangleq2\pi f_c$ over the probing band.

Define Green's function as
\begin{equation}
G_k(\mathbf r,\mathbf r')\triangleq \frac{-\mathrm{j}}{4}\kappa^2(\omega_k)\,
H_0^{(2)}\!\bigl(\kappa(\omega_k)\|\mathbf r-\mathbf r'\|\bigr),
\label{eq:green_kernel}
\end{equation}
where $H_0^{(2)}$ is the zeroth-order Hankel function of the second kind, $\kappa(\omega_k)\triangleq \omega_k/C$, and $C$ is the speed of light.

Discretizing $\mathcal{D}$ into $\mathcal{D'}$ yields the contrast vector $\boldsymbol\chi\in\mathbb C^{ M \times 1}$ and the in-domain Green's matrix $\mathbf G_k^{\mathcal D'}\in\mathbb C^{ M\times  M}$.
Let $\mathbf P_k\in\mathbb C^{L\times  M}$ denote the free-space propagation operator between the aperture points on $\Gamma$ and the $  M$ grid points in $\mathcal{D'}$.
Under the co-located assumption, the transmit-to-domain and domain-to-receive operators satisfy
\begin{equation}
\mathbf H_{1,k}=\mathbf P_k^{\mathsf T},\qquad \mathbf H_{2,k}=\mathbf P_k .
\label{eq:colocated_H12}
\end{equation}

Let $\mathbf E_k^{t, \mathcal D'}, \mathbf E_k^{i, \mathcal D'}\in\mathbb C^{M\times 1}$ denote the discretized total and incident fields in $\mathcal{D'}$, respectively.
Then, the discretized Lippmann–Schwinger equation is given by
\begin{equation}
\mathbf E_k^{t, \mathcal D'}
=
\mathbf E_k^{i, \mathcal D'}
+
\mathbf G_k^{\mathcal D'}\diag(\boldsymbol\chi)\mathbf E_k^{t, \mathcal D'}.
\label{eq:ls_matrix}
\end{equation}
Rearranging yields
\begin{equation}
\mathbf E_k^{t, \mathcal D'}
=
\Bigl(\mathbf I_{M}-\mathbf G_k^{\mathcal D'}\diag(\boldsymbol\chi)\Bigr)^{-1}\mathbf E_k^{i, \mathcal D'}.
\label{eq:Et_solution}
\end{equation}
Define the induced (equivalent) scattering source in $\mathcal{D'}$ as
\begin{equation}
\mathbf s_k^{\mathcal D'}\triangleq \diag(\boldsymbol\chi)\mathbf E_k^{t, \mathcal D'}.
\label{eq:source_def}
\end{equation}
Substituting the above expression yields
\begin{equation}
\mathbf s_k^{\mathcal D'}
=
\diag(\boldsymbol\chi)
\Bigl(\mathbf I_{M}-\mathbf G_k^{\mathcal D'}\diag(\boldsymbol\chi)\Bigr)^{-1}\mathbf E_k^{i, \mathcal D'}.
\label{eq:source_solution}
\end{equation}

For a given pilot $\mathbf x_k$, the incident field is given by
\begin{equation}
\mathbf E_k^{i, \mathcal D'}=\mathbf H_{1,k}\mathbf x_k=\mathbf P_k^{\mathsf T}\mathbf x_k.
\label{eq:EiD_def}
\end{equation}
The scattered field observed at the aperture is expressed as
\begin{equation}
\begin{split}
\mathbf y_k
&=
\mathbf H_{2,k}\mathbf s_k^{\mathcal D'}+\mathbf n_k \\
&=
\mathbf P_k\,\diag(\boldsymbol\chi)
\Bigl(\mathbf I_{M}-\mathbf G_k^{\mathcal D'}\diag(\boldsymbol\chi)\Bigr)^{-1}
\mathbf P_k^{\mathsf T}\mathbf x_k
+\mathbf n_k,
\end{split}
\label{eq:yk_em}
\end{equation}
which shows the received signal depends nonlinearly on the contrast $\boldsymbol{\chi}$ due to multiple scattering effects.
To enable tractable inference, we reformulate the model into a linear form  with respect to $\boldsymbol{\chi}$ by introducing an equivalent forward operator.
Accordingly, it can be expressed as:
\begin{equation}
\mathbf y_k
=
\mathbf A_k\boldsymbol\chi
+
\mathbf n_k,
\qquad
\mathbf n_k
\sim
\mathcal{CN}\!\left(
\mathbf 0,\,
\sigma^2 \mathbf I_{L}
\right).
\label{eq:yk_linear}
\end{equation}
Let 
$\mathbf u_k \triangleq \Bigl(\mathbf I_{M}-\mathbf G_k^{\mathcal D'}\diag(\boldsymbol\chi)\Bigr)^{-1}\mathbf P_k^{\mathsf T}\mathbf x_k$ and use the identity $\diag(\boldsymbol\chi)\mathbf u_k=\diag(\mathbf u_k)\boldsymbol\chi$.
Then the forward operator is given by
\begin{equation}
\mathbf A_k
\triangleq
\Bigl(
\mathbf x_k^{\mathsf T}\mathbf P_k
\Bigl(\mathbf I_{M}-\mathbf G_k^{\mathcal D'}\diag(\boldsymbol\chi)\Bigr)^{-\mathsf T}
\Bigr)\circ \mathbf P_k,
\label{eq:Ak_colocated}
\end{equation}
where $\mathbf{A}_k \in \mathbb{C}^{L \times  M}$ implicitly depends on $\boldsymbol{\chi}$ through the multiple-scattering interaction.
%

\subsection{Outer Linearization via Born Iterations}

The observation model in \eqref{eq:yk_linear} is nonlinear with respect to the contrast vector due to multiple scattering.
To obtain a tractable inference problem, we adopt an outer Born-iteration scheme~\cite{b27,b30}, which linearizes the forward model around the current estimate.
At the $n$-th outer iteration, the linearized operator is constructed by evaluating the forward operator at the previous estimate $\boldsymbol\chi^{(n-1)}$:
\begin{equation}
\mathbf A_k^{(n)}
\triangleq
\mathbf A_k\big|_{\boldsymbol\chi=\boldsymbol\chi^{(n-1)}} .
\label{eq:Ak_linearized_def}
\end{equation}
This yields the uncertainty-aware linearized observation model
\begin{equation}
\mathbf y_k
\approx
\mathbf A_k^{(n)}\boldsymbol\chi
+
\tilde{\mathbf n}_k^{(n)},
\label{eq:yk_linearized}
\end{equation}
where $\tilde{\mathbf n}_k^{(n)}$ is an effective noise term accounting for both measurement noise and the mismatch caused by the uncertainty in $\boldsymbol\chi^{(n-1)}$.
When posterior uncertainty of $\boldsymbol\chi^{(n-1)}$ is available, the induced uncertainty of $\mathbf A_k^{(n)}$ can be further propagated into an equation-dependent effective noise covariance.

After collecting multi-frequency observations, the joint inverse problem is formed by stacking $\{\mathbf y_k\}$ and $\{\mathbf A_k^{(n)}\}$ across frequencies.
The resulting linear subproblem in \eqref{eq:yk_linearized} is solved by the proposed \ac{GaBP} inference engine~\cite{b17,b18,b18a}, which is scalable to large-scale linear systems.
The outer iterations are repeated until the prescribed convergence criterion is met.

\section{Unified \ac{GaBP}-Based \ac{DT} Reconstruction}
\label{sec:GaBp_unified}

\subsection{Linear-Gaussian Model for Joint \ac{DT} Reconstruction}
\label{subsec:prob_model_priors}

In conventional \ac{EM} inverse scattering, the antenna geometry is assumed to be known, and the inverse problem reduces to estimating the material contrast $\boldsymbol\chi$ under a fixed forward model.
In practical \ac{ISAC} transceivers, however, the array geometry may be mismatched or unknown.
To explicitly capture the dependency, we introduce the array-geometry state $\boldsymbol\theta$, which parameterizes the propagation operator at subcarrier $k$.
Specifically, $\boldsymbol\theta$ collects the radial and angular coordinates of all array elements.
Accordingly, the propagation operator is expressed as $\mathbf P_k(\boldsymbol\theta)\in\CC^{L  \times  M}$.

The forward operator is then written as
\begin{equation}
\mathbf A_k(\boldsymbol\theta,\boldsymbol\chi)
\triangleq
\left(
\mathbf x_k^{\mathsf T}\mathbf P_k(\boldsymbol\theta)
\big(
\mathbf I_{M}-\mathbf G_k^D\diag(\boldsymbol\chi)
\big)^{-\mathsf T}
\right)
\circ
\mathbf P_k(\boldsymbol\theta),
\label{eq:Ak_theta_chi}
\end{equation}
which depends on $\boldsymbol\theta$ through $\mathbf P_k(\boldsymbol\theta)$ and on $\boldsymbol\chi$ through the multiple-scattering interaction.
Thus, the model follows the dependency chain
\begin{equation}
    \boldsymbol\theta \rightarrow \mathbf P_k(\boldsymbol\theta) \rightarrow \mathbf A_k^{(n)}(\boldsymbol\theta) \rightarrow \mathbf y_k.
\end{equation}
For notational simplicity, we focus on a single subcarrier and omit the subcarrier index $k$ in the following.

The inverse problem therefore shifts from estimating $\boldsymbol\chi$ alone to jointly recovering $\boldsymbol\chi$ and $\boldsymbol\theta$.
This results in a bilinear inverse problem that couples scene and system uncertainties~\cite{b22,b23,b24}.

A key consequence of this bilinear coupling is the presence of scaling ambiguity.
For the purpose of ambiguity analysis, let $\mathbf A^{(n)}(\boldsymbol{\theta})$ denote the linearized forward operator induced by $\boldsymbol{\theta}$.
Then, in the noiseless case, there exists a scalar $\beta\neq 0$ such that
\begin{equation}
\mathbf A^{(n)}(\boldsymbol{\theta})\boldsymbol\chi
=
\mathbf A^{(n)}(\boldsymbol{\theta}/\beta)\,(\beta\boldsymbol\chi),
\label{eq:scaling_ambiguity}
\end{equation}
indicating that different parameter pairs related by $\beta$ yield identical measurements.
This ambiguity renders $\boldsymbol\chi$ non-identifiable along the scaling direction and leads to ill-conditioning if not properly regularized.

After outer linearization, we consider the regularized objective
\begin{equation}
\min_{\boldsymbol\chi,\boldsymbol\theta}
\left\|
\mathbf y-\mathbf A^{(n)}(\boldsymbol\theta)\boldsymbol\chi
\right\|_2^2
+
\lambda_\chi\|\boldsymbol\chi\|_2^2
+
\lambda_\theta\|\boldsymbol\theta-\boldsymbol\theta^{(n)}\|_2^2,
\label{eq:tikh_joint_theta}
\end{equation}
where $\boldsymbol\theta^{(n)}$ denotes the geometry state at the $n$-th outer iteration, and $\lambda_\chi$ and $\lambda_\theta$ are the corresponding regularization parameters. 

We reinterpret these quadratic penalties in a Gaussian framework and solve the resulting large-scale linear-Gaussian subproblems using the proposed \ac{GaBP} engine~\cite{b17,b18}.
Although the joint problem in $(\boldsymbol\chi,\boldsymbol\theta)$ is nonconvex, it becomes conditionally Gaussian when either variable block is fixed.
We therefore adopt an alternating scheme consisting of a $\boldsymbol\chi$-step with fixed $\boldsymbol\theta$, and a $\boldsymbol\theta$-step with fixed $\boldsymbol\chi$.

The prior for the contrast within the \ac{ROI} is specified as
\begin{equation}
\chi_m\sim\mathcal{CN}\left(\mu_m^{(0)},\left(1/\alpha_{\chi,m}^{(0)}\right)\right),\quad \forall m \in\mathcal{M},
\label{eq:chi_prior_init_vec}
\end{equation}
where $\boldsymbol\mu^{(0)}\in\CC^{M}$ collects the prior means, $\alpha_{\chi,m}^{(0)}>0$ denotes the prior precision for pixel $m$, and $\mathcal{M}\triangleq\{1,\ldots,M\}$.

The nominal array is initialized as a \ac{UCA} of radius $R_0$, with
\begin{equation}
\phi_\ell^{(0)}\triangleq \frac{2\pi(\ell-1)}{L},
\quad \forall\ell\in\mathcal{L},
\end{equation}
and nominal element radius $ r_\ell^{(0)}\triangleq R_0 $, where $\mathcal{L}\triangleq\left\{1,2,\cdots,L\right\}$. 

At the $n$-th outer iteration, the geometry increment is modeled as
\begin{equation}
\delta\boldsymbol\theta_\ell
\triangleq
[\delta r_\ell,\ \delta\phi_\ell]^{\mathsf T}
\sim
\Normal(\mathbf 0,\mathbf\Sigma_{\theta,\ell}),
\quad
\mathbf\Sigma_{\theta,\ell}\triangleq\diag(\sigma_r^2,\sigma_\phi^2),
\label{eq:delta_theta_prior}
\end{equation}
where $\sigma_r^2,\sigma_\phi^2$ are corresponding variances.

\subsection{\texorpdfstring{$\boldsymbol\chi$-step}{chi-step}: Complex \ac{GaBP} Inference}

At the $n$-th outer iteration, the geometry state $\boldsymbol\theta^{(n-1)}$ is treated as fixed.
Accordingly, we define the linearized scattering operator
$\mathbf A^{(n)}
\triangleq
\mathbf A\!\big(\boldsymbol\theta^{(n-1)}\big)$
and treat it as a known within the current $\boldsymbol\chi$-step.
The $\ell$-th element of the measurement vector $\bm{y}$ is given by
\begin{equation}
y_{\ell}
=
\sum_{m\in \mathcal N(\ell)} A_{\ell,m}\,\chi_m + n_{\ell},
\quad
n_{\ell} \sim \CN(0,\sigma^2),
\label{eq:scalar_meas_model}
\end{equation}
for $\ell\in\mathcal{L}$, where $\ell$ indexes the aperture elements and $m\in\mathcal{M}$ indexes the pixels.
Here, $\mathcal N(\ell)$ denotes the set of variables connected to factor node $\ell$, and $\mathcal N(m)$ denotes the set of factors connected to variable node $m$.


Each contrast variable $\chi_m$ is assigned an independent proper complex Gaussian prior.
At the $n$-th outer iteration, the prior is given by
\begin{equation}
p(\chi_m)
=
\CN\!\big(\chi_m;\,u_m^{(n)},\,v_m^{(n)}\big),
\label{eq:chi_prior_scalar_pdf}
\end{equation}
where $u_m^{(n)}=\mu_m^{(n)}$ and
$v_m^{(n)}= 1/ \alpha_{\chi,m}^{(n)}$ denote the prior mean and variance, respectively.

Expressed in natural-parameter form, the prior becomes
\begin{equation}
p(\chi_m)
\propto
\exp\!\left(
-\tau_{m}^{\rm prior}\,|\chi_m|^2
+ 2\,\Re\!\big\{\left(\eta_{m}^{\rm prior}\right)^*\chi_m\big\}
\right),
\label{eq:chi_prior_natural}
\end{equation}
where
\begin{equation}
\tau_{m}^{\rm prior}
=
\alpha_{\chi,m}^{(n)},
\qquad
\eta_{m}^{\rm prior}
=
\alpha_{\chi,m}^{(n)}\,u_m^{(n)}.
\label{eq:chi_prior_nat_params}
\end{equation}

The resulting factor graph consists of  $M$ variable nodes and $L$ factor nodes, each factor corresponds to a  linear mixing constraint.

\subsubsection{Factor-to-Variable (F2V) Update}

For a factor node $\ell$, assume that incoming \ac{V2F} messages are Gaussian:
\begin{equation}
\label{eq:incoming_v2f_gauss}
\varDelta_{m\to \ell}(\chi_m)=\CN(\chi_m;\mu_{m\to \ell},v_{m\to \ell}),
\quad \forall m\in \mathcal N(\ell),
\end{equation}
where $(\mu_{m\to \ell},v_{m\to \ell})$ denote the mean and variance of the incoming messages, whose Gaussian form will be established in the subsequent \ac{V2F} update. 

The \ac{F2V} message is obtained by marginalizing over all variables except $\chi_m$.
Although the exact marginal is not strictly Gaussian, it can be accurately approximated as Gaussian due to the aggregation of many independent contributions in the linear mixture, which can be justified via a central limit theorem argument in the large-system regime~\cite{b25}.


Define the row-wise prediction
\begin{equation}
\label{eq:row_pred_full}
\widehat y_{\ell} \triangleq \sum_{m\in \mathcal N(\ell)}A_{\ell,m}\mu_{m\to \ell},
\quad
S_{\ell} \triangleq \sum_{m\in \mathcal N(\ell)}|A_{\ell,m}|^2\,v_{m\to \ell}.
\end{equation}
Excluding variable $m$, we have
\begin{equation}
\label{eq:row_pred_loo}
\widehat y_{\ell\setminus m}
\triangleq
\widehat y_{\ell}-A_{\ell,m}\mu_{m\to \ell},
\quad
S_{\ell\setminus m}
\triangleq
S_{\ell}-|A_{\ell,m}|^2\,v_{m\to \ell}.
\end{equation}
Define the effective residual and variance:
\begin{equation}
\label{eq:eff_res_var}
y_{\ell\setminus m}^{e}
\triangleq
y_{\ell}-\widehat y_{\ell\setminus m},
\quad
\sigma_{\ell\setminus m}^2
\triangleq
\sigma^2+S_{\ell\setminus m}.
\end{equation}
The resulting \ac{F2V} message is approximated as
\begin{equation}
\label{eq:f2v_gauss}
\varDelta_{\ell\to m}(\chi_m)
=
\CN\!\left(
\chi_m;\ \frac{y_{\ell\setminus m}^{e}}{A_{\ell,m}},\
\frac{\sigma_{\ell\setminus m}^2}{|A_{\ell,m}|^2}
\right).
\end{equation}
with natural parameters
\begin{equation}
\label{eq:f2v_nat_update}
\tau_{\ell\to m}
\triangleq
\frac{|A_{\ell,m}|^2}{\sigma_{\ell\setminus m}^2},
\quad
\eta_{\ell\to m}
\triangleq
\frac{A_{\ell,m}^{*}\,y_{\ell\setminus m}^{e}}{\sigma_{\ell\setminus m}^2}.
\end{equation}

\subsubsection{Variable-to-Factor (V2F) Update}

Given Gaussian \ac{F2V} messages and a Gaussian prior, the \ac{V2F} update is obtained by combining them:
\begin{equation}
\label{eq:v2f_rule}
\varDelta_{m\to \ell}(\chi_m)
\propto
p(\chi_m)
\prod_{\ell'\in \mathcal N(m)\setminus\{\ell\}} \varDelta_{\ell'\to m}(\chi_m).
\end{equation}
%
%
Since all terms are Gaussian in natural form, the resulting message remains Gaussian, with additive parameters:
\begin{equation}
\label{eq:v2f_nat_sum}
\begin{aligned}
\tau_{m\to \ell}
&=
\tau_{m}^{\rm prior}
+\sum_{\ell'\in \mathcal N(m)\setminus\{\ell\}}\tau_{\ell'\to m}, \\
\eta_{m\to \ell}
&=
\eta_{m}^{\rm prior}
+\sum_{\ell'\in \mathcal N(m)\setminus\{\ell\}}\eta_{\ell'\to m}.
\end{aligned}
\end{equation}
Equivalently, we may express
\begin{equation}
\label{eq:v2f_moments}
\begin{aligned}
\varDelta_{m\to \ell}(\chi_m)
&=
\CN(\chi_m;\mu_{m\to \ell},v_{m\to \ell}), \\
v_{m\to \ell}
&=
\tau_{m\to \ell}^{-1}, \\
\mu_{m\to \ell}
&=
\eta_{m\to \ell}/\tau_{m\to \ell}.
\end{aligned}
\end{equation}

\subsubsection{Final Posterior Estimate}

After a prescribed number of \ac{F2V} and \ac{V2F} message-passing iterations (or upon convergence), the posterior message at $\chi_m$ is given by
\begin{equation}
\label{eq:belief_def}
\varDelta(\chi_m)\ \propto\
p(\chi_m)\prod_{\ell\in \mathcal N(m)} \varDelta_{\ell\to m}(\chi_m).
\end{equation}
which remains Gaussian in natural form:
\begin{equation}
\label{eq:belief_nat_params}
\begin{aligned}
\tau_{m}^{\rm post}
&=
\tau_{m}^{\rm prior}
+\sum_{\ell\in \mathcal N(m)}\tau_{\ell\to m}, \\
\eta_{m}^{\rm post}
&=
\eta_{m}^{\rm prior}
+\sum_{\ell\in \mathcal N(m)}\eta_{\ell\to m}.
\end{aligned}
\end{equation}
Hence the posterior mean and variance are
\begin{equation}
\label{eq:post_stats}
v_m^{\rm post}=(\tau_{m}^{\rm post})^{-1},
\qquad
u_m^{\rm post}=\eta_{m}^{\rm post}/\tau_{m}^{\rm post},
\end{equation}
and we take $\chi_m^{(n)}=u_m^{\rm post}$ as the \ac{GaBP} estimate at the $n$-th outer iteration.

The overall procedure is summarized in Algorithm~1. Under the current parameter setting, the computational complexity of the $\chi$-step is 
$\mathcal{O}\!\left(NI_{\chi}LM\right)$, and $I_{\chi}$ denotes the number of inner \ac{GaBP} iterations.

\begin{algorithm}[t]
\caption{Complex GaBP $\boldsymbol\chi$-step}
\label{alg:chi_step_pseudo}
\begin{algorithmic}[1]
\State \textbf{Input:} $\mathbf y$, $\mathbf A^{(n)}$, $\sigma^2$,
prior $\{(\tau_{m}^{\rm prior},\eta_{m}^{\rm prior})\}$,
$\epsilon_\chi$
\State Initialize $(\tau_{\ell\to m},\eta_{\ell\to m}) \leftarrow 0$,
$\widehat{\boldsymbol\chi}^{\mathrm{old}} \leftarrow \mathbf 0$

\Repeat

  \For{$\forall m \in \mathcal{M}$}
    \State Compute $(\tau_{m\to \ell},\eta_{m\to \ell})$ for all $\ell \in \mathcal N(m)$ via \eqref{eq:v2f_nat_sum}
  \EndFor

  \For{$\forall \ell \in \mathcal{L}$}
    \State Compute $\widehat y_{\ell}$ and $S_{\ell}$ via \eqref{eq:row_pred_full}
    \State Update $(\tau_{\ell\to m},\eta_{\ell\to m})$ for all $m\in\mathcal N(\ell)$ via \eqref{eq:f2v_nat_update}
  \EndFor

  \For{$\forall m \in \mathcal{M}$}
    \State Compute $(\tau_m^{\rm post},\eta_m^{\rm post})$ via \eqref{eq:belief_nat_params}
    \State Compute $(u_m^{\rm post},v_m^{\rm post})$ via \eqref{eq:post_stats}, set $\widehat\chi_m=u_m^{\rm post}$
  \EndFor

  \State Store $\boldsymbol\chi^{(n)}=[\widehat\chi_1,\ldots,\widehat\chi_{M}]^{\mathsf T}$
  \State Compute
  $
  \delta_\chi \leftarrow
  \dfrac{\|\boldsymbol\chi^{(n)}-\widehat{\boldsymbol\chi}^{\mathrm{old}}\|_2}
  {\max(\|\widehat{\boldsymbol\chi}^{\mathrm{old}}\|_2,10^{-8})}
  $
  \State Update $\widehat{\boldsymbol\chi}^{\mathrm{old}} \leftarrow \boldsymbol\chi^{(n)}$

\Until{$\delta_\chi \le \epsilon_\chi$}

\State \textbf{Output:} $\boldsymbol\chi^{(n)}$, $\{v_m^{\rm post}\}$
\end{algorithmic}
\end{algorithm}

\subsection{\texorpdfstring{$\boldsymbol{\theta}$-step}{theta-step}: Uncertainty-Aware Array Geometry Update}
\label{subsec:theta_step}

In the following, we use the subscript $l \in \mathcal{L}$ to index the observation equations, while $\ell$ indexes the antenna elements.
Although these indices coincide numerically in the single-subcarrier case, we distinguish them to emphasize that each antenna parameter $\theta_\ell$ influences all observations $y_l$.
%
This distinction becomes essential in the Jacobian representation introduced below.

Using the posterior variances $\{v_m^{\rm post}\}_{m=1}^{M}$ obtained from the $\boldsymbol\chi$-step, we first propagate the uncertainty of
$\boldsymbol\chi^{(n)}$ into the observation domain.
For each observation index $l$, let $\mathbf A_{l,:}^{(n)}\in\mathbb C^{1\times M}$ denote the corresponding row of the linearized scattering operator.
The resulting variance contribution induced by $\boldsymbol\chi$ is defined as
\begin{equation}
\label{eq:nu_def_allvars}
\nu_l^{(n)}
\triangleq
\sum_{m=1}^{M}
\bigl|\mathbf A_{l,m}^{(n)}\bigr|^2\,v_m^{\rm post}.
\end{equation}

We now describe the $\boldsymbol{\theta}$-step under the same Gaussian modeling framework.
The geometry state is parameterized as
\begin{equation}
\boldsymbol\theta
=
\{(r_\ell,\phi_\ell)\}_{\ell=1}^{L}
\in\mathbb R^{2L}.
\end{equation}

At the $n$-th outer iteration (totally $N$ iterations), fixing $\boldsymbol\chi^{(n)}$,
we define the residual under the current reference geometry state
$\boldsymbol\theta^{(n-1)}$ as
\begin{equation}
\label{eq:astep_residual_def}
\mathbf e
\triangleq
\mathbf y-
\widehat{\mathbf y}\!\left(
\boldsymbol\theta^{(n-1)};\boldsymbol\chi^{(n)}
\right),
\end{equation}
where $\widehat{\mathbf y}(\boldsymbol\theta^{(n-1)};\boldsymbol\chi^{(n)})
\triangleq
\mathbf A(\boldsymbol\theta^{(n-1)};\boldsymbol\chi^{(n)})\boldsymbol\chi^{(n)}$.

Let $\delta\boldsymbol\theta 
\triangleq
\big[
\delta\boldsymbol\theta_1^{\mathsf T},
\dots,
\delta\boldsymbol\theta_{L}^{\mathsf T}
\big]^{\mathsf T}
\in\mathbb R^{2L}$ denote the stacked geometry increment, with per-antenna block
\begin{equation}
\delta\boldsymbol\theta_\ell
\triangleq
\begin{bmatrix}
\delta r_\ell\\[2pt]
\delta\phi_\ell
\end{bmatrix}
\in\mathbb R^2.
\label{eq:delta_theta_global_def}
\end{equation}
Linearizing the forward model with respect to $\boldsymbol\theta$ around
$\boldsymbol\theta^{(n-1)}$ yields
\begin{equation}
\label{eq:theta_global_linear_model}
\mathbf e
\approx
\mathbf J\,\delta\boldsymbol\theta+\boldsymbol\varepsilon,
\end{equation}
where
$\mathbf J\in\mathbb C^{L\times 2L}$ is the stacked Jacobian matrix and $\boldsymbol\varepsilon$ represents the effective model error.

For each observation $l$, the effective noise is modeled as
\begin{equation}
\label{eq:ve_def_global}
\varepsilon_l \triangleq [\boldsymbol\varepsilon]_l
\sim \CN\!\left(0,v_{e,l}^{(n)}\right),
\quad
v_{e,l}^{(n)}
\triangleq
\sigma^2+\nu_l^{(n)}+\xi_l^{(n-1)}.
\end{equation}
Here, $[\boldsymbol\varepsilon]_l$ denotes the $l$-th entry of the vector $\boldsymbol\varepsilon$, $\nu_l^{(n)}$ accounts for propagated contrast uncertainty, and $\xi_l^{(n-1)}\ge 0$ accounts for geometry-induced uncertainty carried over from the previous iteration.
These contributions are treated as independent and combined additively to form the effective noise model.

\begin{algorithm}[t]
\caption{Approximate Gaussian Fusion for the \texorpdfstring{$\boldsymbol{\theta}$-step}{theta-step}}
\label{alg:astep_pseudo}
\begin{algorithmic}[1]
\State \textbf{Input:} $\mathbf y$, $\boldsymbol\theta^{(n-1)}$, $\boldsymbol\chi^{(n)}$,
$\{v_m^{\rm post}\}_{m=1}^{M}$, $\sigma^2$,
$\{\mathbf\Sigma_{\theta,\ell}\}_{\ell=1}^{L}$, $\gamma$

\State \textbf{Output:} $\boldsymbol\theta^{(n)}$,
$\{\mathbf\Sigma_{\delta\theta_\ell}^{\rm post}\}_{\ell=1}^{L}$,
and updated $\{\xi_l^{(n)}\}_{l=1}^{L}$

\State Compute $\nu_l^{(n)}$ for all $l$ via \eqref{eq:nu_def_allvars}

\State Compute $v_{e,l}^{(n)}=\sigma^2+\nu_l^{(n)}+\xi_l^{(n-1)}$, set $w_l^{(n)}=(v_{e,l}^{(n)})^{-1}$

\State $\mathbf e \leftarrow \mathbf y-\widehat{\mathbf y}(\boldsymbol\theta^{(n-1)};\boldsymbol\chi^{(n)})$

\For{$\forall \ell \in \mathcal{L}$}
    
    \State Initialize $\mathbf \Lambda_\ell^{\rm post}\leftarrow \mathbf\Sigma_{\theta,\ell}^{-1}$ and $\mathbf h_\ell^{\rm post}\leftarrow \mathbf 0$
    \For{$\forall l \in \mathcal{L}$}
        \State Compute $\mathbf\Lambda_{\ell,l}$ and $\mathbf b_{\ell,l}$ via \eqref{eq:Lambda_m_def_allvars}
        \State $\mathbf \Lambda_\ell^{\rm post}\leftarrow \mathbf\Lambda_\ell^{\rm post}+2\mathbf\Lambda_{\ell,l}$
        \State $\mathbf h_\ell^{\rm post}\leftarrow \mathbf h_\ell^{\rm post}+\mathbf b_{\ell,l}$
    \EndFor
    \State $\mathbf\Sigma_{\delta\theta_\ell}^{\rm post}\leftarrow (\mathbf \Lambda_\ell^{\rm post})^{-1}$
    \State $\widehat{\delta\boldsymbol\theta}_\ell\leftarrow \mathbf\Sigma_{\delta\theta_\ell}^{\rm post}\mathbf h_\ell^{\rm post}$
    \State Update $\boldsymbol\theta_\ell^{(n)}$, $\xi_l^{(n)}$ via (\ref{eq:theta_update2}) and (\ref{eq:xi_def_allvars})
 \EndFor

\end{algorithmic}
\end{algorithm}

Define the precision weights as
\begin{equation}
\label{eq:weight_def}
w_l^{(n)}
\triangleq
\big(v_{e,l}^{(n)}\big)^{-1},
\quad
\mathbf W^{(n)}
\triangleq
\diag\!\big(w_1^{(n)},\dots,w_L^{(n)}\big).
\end{equation}
The model can then be written in whitened form as
\begin{equation}
\label{eq:theta_global_whitened}
\big(\mathbf W^{(n)}\big)^{1/2}\mathbf e
=
\big(\mathbf W^{(n)}\big)^{1/2}\mathbf J\,\delta\boldsymbol\theta
+
\tilde{\boldsymbol\varepsilon},
\end{equation}
which defines a global linear-Gaussian calibration problem, with $\tilde{\boldsymbol\varepsilon}\triangleq (\mathbf W^{(n)})^{1/2}\boldsymbol\varepsilon$ denote the whitened effective noise.

Assuming independent Gaussian priors across antenna blocks, we model the geometry increments as
\begin{equation}
\label{eq:theta_prior_mlocks}
\delta\boldsymbol\theta_\ell
\sim
\mathcal N(\mathbf 0,\mathbf\Sigma_{\theta,\ell}),
\qquad
\ell=1,\dots,L,
\end{equation}
where $\mathbf\Sigma_{\theta,\ell}$ denotes the covariance matrix associated with antenna $\ell$.

Combining the linearized observation model with the Gaussian prior yields a global linear-Gaussian estimation problem.
The corresponding \ac{MAP} estimator is given by
\begin{equation}
\label{eq:theta_global_map}
\min_{\delta\boldsymbol\theta}
\left\|
\big(\mathbf W^{(n)}\big)^{1/2}
\big(\mathbf e-\mathbf J\delta\boldsymbol\theta\big)
\right\|_2^2
+
\sum_{\ell=1}^{L}
\delta\boldsymbol\theta_\ell^{\mathsf T}
\mathbf\Sigma_{\theta,\ell}^{-1}
\delta\boldsymbol\theta_\ell.
\end{equation}
This corresponds to Gaussian inference on a global factor graph induced by the linearized model.

However, exact inference is computationally prohibitive due to dense cross-block couplings in the Jacobian, which induce strong dependencies across antenna parameters.

To obtain a tractable solution, we adopt a block-diagonal approximation of the global information, thereby decoupling the geometry increments across antennas and leading to a set of per-antenna subproblems.

For each antenna $\ell$, let $\mathbf J_\ell\in\mathbb C^{L\times 2}$ denote the Jacobian block associated with the two geometric parameters of antenna $\ell$. Thus,
\begin{equation}
\label{eq:J_l_def}
\mathbf J
=
\big[
\mathbf J_1,\,
\mathbf J_2,\,
\ldots,\,
\mathbf J_L
\big],
\qquad
\mathbf J_\ell\in\mathbb C^{L\times 2}.
\end{equation}
The $l$-th row of $\mathbf J_\ell$ is denoted by
\begin{equation}
\label{eq:jacobian_row_def}
\mathbf j_{l,\ell}
\triangleq
\mathbf J_\ell(l,:)
=
\left[
\frac{\partial\,\widehat y_l(\boldsymbol\theta;\boldsymbol\chi^{(n)})}{\partial r_\ell},
\;
\frac{\partial\,\widehat y_l(\boldsymbol\theta;\boldsymbol\chi^{(n)})}{\partial \phi_\ell}
\right]
\in\mathbb C^{1\times 2},
\end{equation}
which characterizes the sensitivity of the $l$-th observation equation to the two geometric parameters of antenna $\ell$.

Under this approximation, the global model decomposes into per-antenna local models. In particular, the contribution of antenna $\ell$ to the $l$-th observation is expressed as
\begin{equation}
\label{eq:theta_local_model}
e_l \triangleq \mathbf {[e]}_l
\approx
\mathbf j_{l,\ell}\,\delta\boldsymbol\theta_\ell
+
\varepsilon_l,
\qquad \forall l\in\mathcal{L},
\end{equation}
with equation-dependent precision $w_l^{(n)}$ inherited from the global model.

Accordingly, each scalar residual equation induces a unary Gaussian factor
\begin{equation}
\label{eq:theta_local_factor}
f_{\ell,l}(\delta\boldsymbol\theta_\ell)
\propto
\exp\!\left(
-w_l^{(n)}
\big|e_l-\mathbf j_{l,\ell}\delta\boldsymbol\theta_\ell\big|^2
\right).
\end{equation}
Expanding the quadratic form yields
\begin{equation}
\label{eq:theta_local_info_form}
f_{\ell,l}(\delta\boldsymbol\theta_\ell)
\propto
\exp\!\left(
-\delta\boldsymbol\theta_\ell^{\mathsf T}\mathbf\Lambda_{\ell,l}\delta\boldsymbol\theta_\ell
+
\mathbf b_{\ell,l}^{\mathsf T}\delta\boldsymbol\theta_\ell
\right),
\end{equation}
where

\begin{equation}
\label{eq:Lambda_m_def_allvars}
\mathbf\Lambda_{\ell,l}
\triangleq
\Re\!\left(
w_l^{(n)}\mathbf j_{l,\ell}^{\mathsf H}\mathbf j_{l,\ell}
\right)
\in\mathbb R^{2\times 2},
\qquad
\mathbf b_{\ell,l}
\triangleq
2\,\Re\!\left(
w_l^{(n)}\mathbf j_{l,\ell}^{\mathsf H}e_l
\right)
\in\mathbb R^{2}.
\end{equation}

Combining the Gaussian prior in \eqref{eq:theta_prior_mlocks} with all local factors associated with antenna $\ell$ yields the approximate posterior
\begin{equation}
\label{eq:theta_local_posterior}
p(\delta\boldsymbol\theta_\ell\mid \mathbf e)
\propto
\exp\!\left(
-\frac12\delta\boldsymbol\theta_\ell^{\mathsf T}\mathbf\Lambda_\ell^{\rm post}\delta\boldsymbol\theta_\ell
+
(\mathbf h_\ell^{\rm post})^{\mathsf T}\delta\boldsymbol\theta_\ell
\right),
\end{equation}
where
\begin{equation}
\mathbf\Lambda_\ell^{\rm post}
=
\mathbf\Sigma_{\theta,\ell}^{-1}
+
2\sum_{l=1}^{L}\mathbf\Lambda_{\ell,l},
\qquad
\mathbf h_\ell^{\rm post}
=
\sum_{l=1}^{L}\mathbf b_{\ell,l}.
\end{equation}

Thus, the posterior covariance and mean are given by
\begin{equation}
\label{eq:theta_local_moments}
\mathbf\Sigma_{\delta\theta_\ell}^{\rm post}
=
(\mathbf\Lambda_\ell^{\rm post})^{-1},
\qquad
\widehat{\delta\boldsymbol\theta}_\ell
=
\mathbf\Sigma_{\delta\theta_\ell}^{\rm post}\mathbf h_\ell^{\rm post}.
\end{equation}

The geometry state is then updated blockwise as
\begin{equation}
\label{eq:theta_update2}
\boldsymbol\theta_\ell^{(n)}
=
\boldsymbol\theta_\ell^{(n-1)}
+
\gamma\,\widehat{\delta\boldsymbol\theta}_\ell,
\qquad
\forall \ell\in\mathcal{L},
\end{equation}
where $\gamma\in(0,1]$ is a damping factor.

Finally, the posterior covariance is propagated back to the observation domain to quantify the residual uncertainty due to geometry estimation:
\begin{equation}
\label{eq:xi_def_allvars}
\xi_l^{(n)}
\triangleq
\sum_{\ell=1}^{L}
\mathbf j_{l,\ell}\,
\mathbf\Sigma_{\delta\theta_\ell}^{\rm post}\,
\mathbf j_{l,\ell}^{\mathsf H}.
\end{equation}

The overall procedure is summarized in Algorithm~\ref{alg:astep_pseudo}. Under the current parameter setting, the computational complexity of the $\theta$-step is $\mathcal{O}(NLM+NL^2)$, which can be approximated as $\mathcal{O}(NLM)$ since the $M$ is typically much larger than $L$

\section{\ac{DT}-Based Closed-Loop Beam Management}

\subsection{DT-Guided Codebook Preselection}

Based on the proposed \ac{GaBP} framework, we jointly reconstruct the \ac{EM} properties of environmental scatterers and the antenna-array geometry, enabling simultaneous scatter reconstruction and array calibration.
The resulting physical \ac{DT} is incorporated into the system design and control loop as a scene-specific prior for the propagation model.

Leveraging this prior, the system supports scenario-adaptive parameter tuning and resource reconfiguration within a closed-loop \textbf{sensing--modeling--control--feedback} framework.
Here, we first consider codebook preselection to narrow the candidate beam set before beam optimization.

In conventional decoupled systems, identifying the optimal transmit beam requires exhaustive beam training~\cite{b32}, in which all codebook beams are sequentially probed.
In large-scale \ac{MIMO} systems, this incurs substantial overhead in training time, signaling cost, and computational complexity.
By contrast, with \ac{DT}-derived propagation priors, dominant departure directions can be inferred directly from the reconstructed scattering geometry, enabling targeted beam transmission without exhaustive codebook sweeping.
This significantly reduces training overhead and search complexity.

Consider a co-located sensing setup in which a \ac{UCA} performs both transmission and reception of multi-frequency probing signals.
Based on the resulting observations, the array geometry is calibrated via the proposed \ac{GaBP}-based $\boldsymbol{\theta}$-step.
After calibration, the \ac{UCA} is used as the receive array.
For the subsequent beam-control design, we introduce a separate transmit \ac{ULA} with $N_{\mathrm{TX}}$ antenna elements and half-wavelength inter-element spacing.

The $\boldsymbol\chi$-step of \ac{GaBP} yields a discretized contrast distribution over the sensing region $\mathcal{D}$.
Suppose that the reconstruction in $\mathcal{D}$ results in $Q$ scatterer clusters $\{\Omega_q\}_{q=1}^{Q}$, within each of which the contrast is approximately uniform.
We use only the real part of $\chi$ to compute the contrast energy, since $\Re\{\chi\}$ captures the permittivity contribution that dominates the stored electric energy and directly characterizes dielectric contrast, whereas $\Im\{\chi\}$ is primarily associated with conductive loss and is more sensitive to modeling mismatch and measurement noise.

According to \eqref{eq:contrast}, the relative permittivity is given by
$
\varepsilon_r(\mathbf r)=\Re\{\chi(\mathbf r)\}+1.
$
We then define the contrast energy density of the $q$-th scatterer cluster as
\begin{equation}
\mathcal{E}(\mathbf r)\triangleq \left|\varepsilon_r(\mathbf r)\right|^2,
\qquad \mathbf r\in\Omega_q.
\end{equation}

To handle multiple reconstructed scatterer clusters, we compute an energy-weighted centroid for each cluster and the corresponding geometric departure angle.
Specifically, the scattering center of the $q$-th cluster is defined as
\begin{equation}
\mathbf r^{\mathrm{scat}}_q
\triangleq
\frac{\displaystyle\sum_{\mathbf r_q \in \Omega_q} 
\mathcal{E}(\mathbf r_q)\,\mathbf r_q}
{\displaystyle\sum_{\mathbf r_q \in \Omega_q} 
\mathcal{E}(\mathbf r_q)},
\quad \forall q\in\mathcal{Q},
\end{equation}
and the associated cluster energy weight is defined as
\begin{equation}
w_q
\triangleq
\sum_{\mathbf r_q \in \Omega_q}\mathcal{E}(\mathbf r_q),
\quad \forall q\in\mathcal{Q},
\end{equation}
where $\mathcal{Q}\triangleq \left\{1,2,\cdots, Q\right\}$, and $ \mathbf{r}_q $ is a pixel in cluster $\Omega_q$.

\begin{figure}[!t]
  \centering
  \includegraphics[width=1\columnwidth]{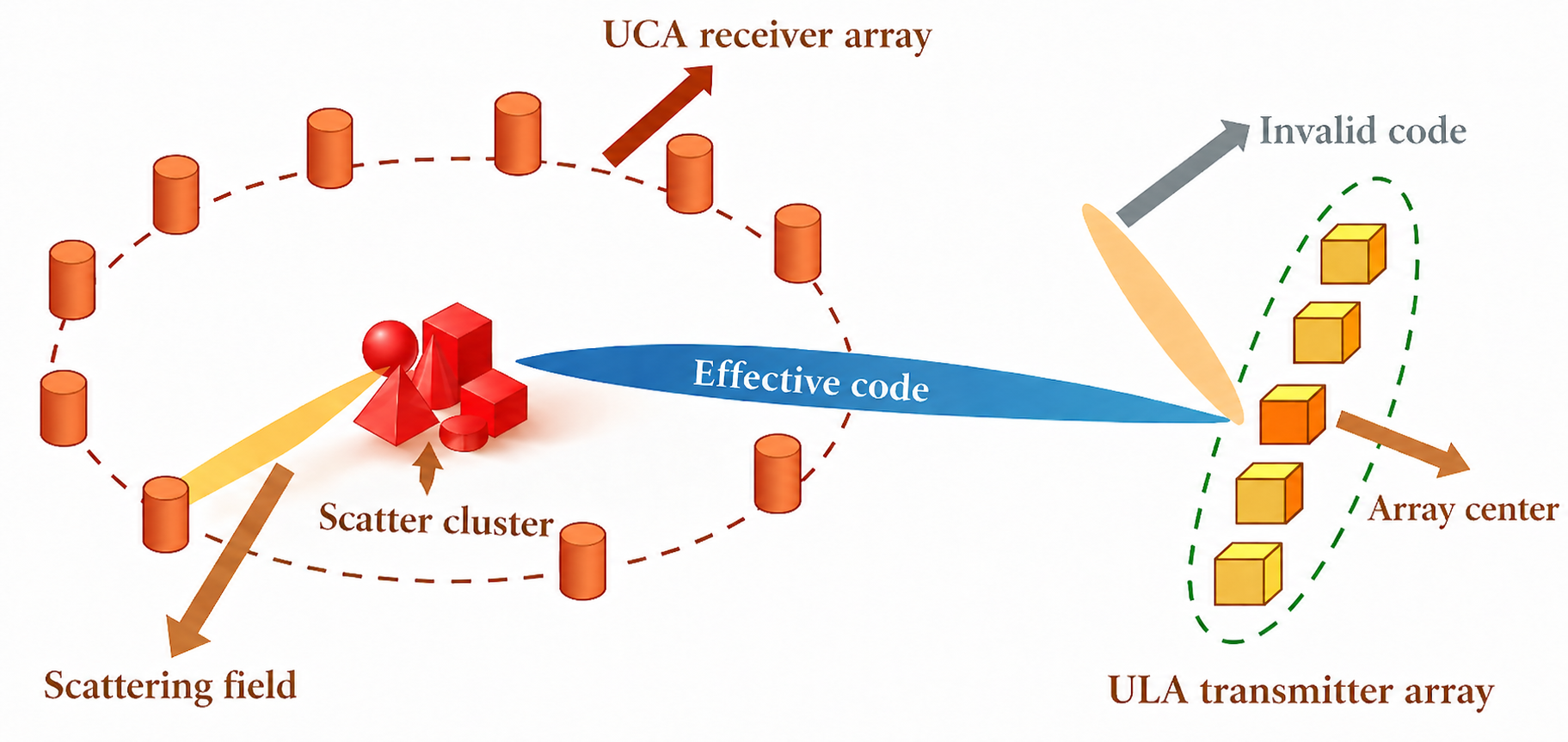}
  \caption{Scatter cluster prior-guided codebook preselection.}
  \label{fig:code_preselection}
\end{figure}

Let $\mathbf o_{\mathrm{TX}}\triangleq[x_{\mathrm{TX}},\,y_{\mathrm{TX}}]^{\mathsf T}$ denote the geometric center of the \ac{TX} array, and let $\mathbf r^{\mathrm{scat}}_q \triangleq[x^{\mathrm{scat}}_{q},\,y^{\mathrm{scat}}_{q}]^{\mathsf T}$.
The geometric departure angle associated with the $q$-th cluster is then given by
\begin{equation}
\varphi_q^{\mathrm{TX}}
\triangleq
\arg\!\left( (x^{\mathrm{scat}}_{q}-x_{\mathrm{TX}})
+ j (y^{\mathrm{scat}}_{q}-y_{\mathrm{TX}}) \right),
\ q \in \mathcal Q,
\end{equation}
where $\arg(\cdot)$ yields an angle of the complex number in $[0,2\pi)$.

Accordingly, the first step of the closed-loop beam-control pipeline returns a set of \ac{DT}-inferred departure angles by selecting the $Z$ most energetic clusters:
\begin{equation}
\mathcal Q_{\mathrm{sel}}
\triangleq
\operatorname{Top}\!-\!Z\big(\{w_q\}_{q=1}^{Q}\big),
\qquad
\Phi_{\mathrm{DT}}
\triangleq
\{\varphi_q^{\mathrm{TX}}:\, q\in\mathcal Q_{\mathrm{sel}}\},
\end{equation}
where $\operatorname{Top}\!-\!Z(\cdot)$ returns the indices of the $Z$ largest weights.

The \ac{DT}-inferred departure-angle set $\Phi_{\mathrm{DT}}$ provides a compact abstraction of the reconstructed propagation environment and serves as a natural interface between sensing and control, as shown in Fig.~\ref{fig:code_preselection}.
Based on this propagation-aware representation, the closed-loop controller first performs a preprocessing step that uses $\Phi_{\mathrm{DT}}$ to guide codebook-based beam preselection.

Consider a \ac{TX} \ac{ULA} with $N_{\mathrm{TX}}$ elements and half-wavelength spacing.
The corresponding steering vector is defined as~\cite{b34}
\begin{equation}
\mathbf a_{\mathrm{TX}}(\varphi)
\triangleq
\frac{1}{\sqrt{N_{\mathrm{TX}}}}
\begin{bmatrix}
e^{-j\frac{2\pi}{\lambda_c} d\, g_{0}\sin\varphi}\\
\vdots\\
e^{-j\frac{2\pi}{\lambda_c} d\, g_{(N_{\mathrm{TX}}-1)}\sin\varphi}
\end{bmatrix},
\end{equation}
where $\lambda_c= C/f_c $ is central wavelength, $d=\lambda_c/2$ is the inter-element spacing, and $g_v$ denotes the centered element index given by
\begin{equation}
g_v \triangleq v-\frac{N_{\mathrm{TX}}-1}{2},\qquad v=0,1,\ldots,N_{\mathrm{TX}}-1.
\end{equation}

We define a predefined angular codebook consisting of $U$ uniformly spaced azimuth directions over $[0,2\pi)$ as
\begin{equation}
\Phi_{\mathrm{cb}}
\triangleq
\left\{
\varphi_u \;\middle|\;
\varphi_u = \frac{2\pi(u-1)}{U},\;
u=1,2,\ldots,U
\right\},
\end{equation}
where these angles represent geometric directions of departure, rather than the intrinsic steering angle domain of the array.

The \ac{TX} \ac{BF} codebook is then constructed as
\begin{equation}
\mathbf F_{\mathrm{TX}}
\triangleq
\left[
\mathbf a_{\mathrm{TX}}(\varphi_1),\,
\mathbf a_{\mathrm{TX}}(\varphi_2),\,
\ldots,\,
\mathbf a_{\mathrm{TX}}(\varphi_U)
\right].
\end{equation}

For each codebook beam, we evaluate its matching score with the \ac{DT}-inferred departure-angle set $\Phi_{\mathrm{DT}}$ as
\begin{equation}
S_u \triangleq \max_{\varphi'\in\Phi_{\mathrm{DT}}}
\left|\mathbf a_{\mathrm{TX}}^{\mathsf H}(\varphi_u)\,\mathbf a_{\mathrm{TX}}(\varphi')\right|^{2},
\quad u=1,2,\ldots,U,
\end{equation}
where $\varphi_u\in\Phi_{\mathrm{cb}}$ denotes the $u$-th codebook direction. 

We then select $B$ beams with the largest scores, i.e.,
\begin{equation}
\label{eq:beamselect}
\mathcal{B}_{\mathrm{sel}}
\triangleq
\operatorname*{arg\,max}_{\substack{\mathcal{B}\subseteq\{1,\ldots,U\}\\ |\mathcal{B}|=B}}
\sum_{u\in\mathcal{B}} S_u,
\end{equation}
where $\mathcal{B}_{\mathrm{sel}}$ denotes the index set of the selected beams.

This completes the first technical step of the proposed closed-loop beam-control framework, namely the identification of candidate beam directions for subsequent robust beam design.

\subsection{Closed-Loop \ac{BF}}

The second technical step of the proposed closed-loop beam-control framework is to perform robust downlink \ac{BF} based on the candidate beams identified in the preceding codebook-preselection step.
The purpose of this step is not to introduce a new \ac{BF} algorithm, but rather to provide a principled way of exploiting the \ac{DT}-inferred cluster parameters to compute a \ac{TX} beam under a limited on-air training budget.

With the dominant transmission directions explicitly characterized by the preceding codebook-preselection step,
the \ac{BF} design can be restricted to a low-dimensional subspace spanned by
a subset of candidate beams.
Specifically, let $\mathcal{B}_{\mathrm{sel}}$ denote the index set of the shortlisted codewords, and collect the corresponding codewords into the matrix
\begin{equation}
\mathbf W_{\mathrm{DT}} \triangleq \left[\mathbf a_{\mathrm{TX}}(\varphi_u)\right]_{u\in\mathcal{B}_{\mathrm{sel}}}.
\end{equation}
The \ac{TX} beamformer is then parameterized as a linear combination within this
\ac{DT}-informed subspace:
\begin{equation}
\mathbf f = \mathbf W_{\mathrm{DT}} \boldsymbol{\eta}, \qquad \|\mathbf f\|_2 = 1,
\label{eq:bf_subspace}
\end{equation}
where $\boldsymbol{\eta}$ is a low-dimensional coefficient vector.

Based on the reconstructed scattering clusters, we construct a frequency-dependent predicted \ac{MIMO} channel as
\begin{equation}
\widehat{\mathbf H}(f) \approx
\sum_{q=1}^{Q}
\alpha_q\,
e^{-j2\pi f \tau_q}\,
\mathbf a_{\mathrm{RX}}(\varphi^{\mathrm{RX}}_q)\,
\mathbf a_{\mathrm{TX}}^{\mathsf H}(\varphi^{\mathrm{TX}}_q),
\label{eq:dt_channel}
\end{equation}
where $\varphi^{\mathrm{RX}}_q$ and $\varphi^{\mathrm{TX}}_q$ denote the azimuth \ac{AoA} and \ac{AoD} extracted from the reconstructed propagation geometry.
Moreover,
$
\tau_q \triangleq \frac{L_q}{C}
$
denotes the geometric delay of the $q$-th cluster, where $L_q$ is the corresponding path length to the scattering center.
The cluster-wise gain proxy is given by
\begin{equation}
\alpha_q \propto \sum_{m:\mathbf r_m\in\Omega_q} \left|[\widehat{\boldsymbol\chi}]_m\right|,
\end{equation}
where $\widehat{\boldsymbol\chi}$ denotes the final contrast estimate.

For notational simplicity, the frequency argument is omitted below when no confusion arises.

The \ac{RX} steering vector is computed in a coordinate-based manner from the calibrated \ac{RX} array geometry returned by the $\boldsymbol\theta$-step.

Let $\mathbf o_{\mathrm{RX}}$ denote the \ac{RX} array center and
$\mathbf r_{\mathrm{RX},\ell}$ be the calibrated coordinate of the $\ell$-th element. Then,
\begin{equation}
\begin{aligned}
\mathbf a_{\mathrm{RX}}(\varphi)
&\triangleq
\frac{1}{\sqrt{L}}
\begin{bmatrix}
e^{-j\frac{2\pi f_c}{C}\,(\mathbf r_{\mathrm{RX},1}-\mathbf o_{\mathrm{RX}})^{\mathsf T}\mathbf u(\varphi)}\\
\vdots\\
e^{-j\frac{2\pi f_c}{C}\,(\mathbf r_{\mathrm{RX},L}-\mathbf o_{\mathrm{RX}})^{\mathsf T}\mathbf u(\varphi)}
\end{bmatrix},\\
\mathbf u(\varphi)
&\triangleq[\cos\varphi,\ \sin\varphi]^{\mathsf T}.
\end{aligned}
\label{eq:aRX_def}
\end{equation}

Assuming maximum-ratio combining (MRC) at the receiver, the predicted received \ac{SNR} at frequency $f$ is
\begin{equation}
\widehat{\gamma}(f;\mathbf f)\triangleq
\frac{\left\|
\widehat{\mathbf H}(f)\,\mathbf f
\right\|_2^2}{\sigma_{\mathrm c}^2},
\end{equation}
and the corresponding wideband \ac{SNR} surrogate is
\begin{equation}
\widehat{\Gamma}(\mathbf f)\triangleq
\sum_{f\in\mathcal F}\widehat{\gamma}(f;\mathbf f)
=
\frac{1}{\sigma_{\mathrm c}^2}\sum_{f\in\mathcal F}
\left\|
\widehat{\mathbf H}(f)\,\mathbf f
\right\|_2^2,
\label{eq:snr_surrogate}
\end{equation}
where $\mathcal F$ denotes the set of considered frequency tones, and $\sigma_{\mathrm c}^2$ denotes the communication-noise power at the receiver.

Since $\widehat{\mathbf H}(f)$ is constructed from \ac{DT}-estimated cluster parameters and calibrated geometry, $\widehat{\Gamma}(\mathbf f)$ inherits \ac{DT} mismatch.
We therefore adopt an outage-aware design~\cite{b35}:
\begin{equation}
\Pr\left\{\widehat{\Gamma}(\mathbf f)\le \Gamma_{\mathrm{th}}\right\}\le \epsilon,
\label{eq:chance_constraint}
\end{equation}
where $\Gamma_{\mathrm{th}}$ denotes the target \ac{SNR} threshold and $\epsilon$ is the allowable outage probability.
Using a standard conservative sufficient condition based on one-sided concentration bounds, we obtain
\begin{equation}
\mu_{\Gamma}(\mathbf f)-\zeta_{\epsilon}\,\sigma_{\Gamma}(\mathbf f)\ \ge\ \Gamma_{\mathrm{th}},
\label{eq:mean_std_mound}
\end{equation}
where $\mu_{\Gamma}(\mathbf f)\triangleq \mathbb E[\widehat{\Gamma}(\mathbf f)]$ and
$\sigma_{\Gamma}^2(\mathbf f)\triangleq \mathrm{Var}[\widehat{\Gamma}(\mathbf f)]$ are taken with respect to the \ac{DT}-induced uncertainty.
Equivalently, we maximize the risk-sensitive utility
\begin{equation}
\max_{\|\mathbf f\|_2=1}\ \mu_{\Gamma}(\mathbf f)-\zeta_{\epsilon}\,\sigma_{\Gamma}(\mathbf f).
\label{eq:rs_objective}
\end{equation}

Substituting $\mathbf f=\mathbf W_{\mathrm{DT}}\boldsymbol{\eta}$ yields a low-dimensional problem over the \ac{DT}-informed subspace.
Since $\mathbf W_{\mathrm{DT}}$ is not necessarily column-orthonormal, we approximate the mean and variance terms by quadratic forms of $\boldsymbol{\eta}$ while retaining a closed-form generalized eigen-solution.
Define
\begin{equation}
\begin{aligned}
\mathbf G_{\mu}
&\triangleq
\frac{1}{\sigma_c^2}\sum_{f\in\mathcal F}
\mathbf W_{\mathrm{DT}}^{\mathsf H}
\mathbf T_{\mu}(f)\,
\mathbf W_{\mathrm{DT}},\ 
\mathbf T_{\mu}(f)\triangleq
\mathbb E\!\left[\widehat{\mathbf H}^{\mathsf H}(f)\widehat{\mathbf H}(f)\right],\\
\mathbf G_{\sigma}
&\triangleq
\frac{1}{\sigma_c^2}\sum_{f\in\mathcal F}
\mathbf W_{\mathrm{DT}}^{\mathsf H}
\mathbf T_{\sigma}(f)\,
\mathbf W_{\mathrm{DT}}.
\end{aligned}
\label{eq:GmuGsig}
\end{equation}
where $\mathbf T_{\sigma}(f)\succeq \mathbf 0$ captures the \ac{DT}-induced uncertainty in the Gram domain.
Importantly, the proposed \ac{GaBP}-based \ac{DT} construction primarily provides posterior mean estimates of the cluster parameters appearing
in \eqref{eq:dt_channel}. 
Let
$
\boldsymbol\psi
\triangleq
\big\{\alpha_q,\tau_q,\varphi_q^{\mathrm{RX}},\varphi_q^{\mathrm{TX}}\big\}_{q=1}^{Q}
$
collect the DT-estimated parameters of the $Q$ dominant clusters, including the complex gain, delay, AoA, and AoD of each cluster.
The covariance matrix $\mathbf C_{\psi}$ quantifies the corresponding estimation uncertainty.
We then obtain $\mathbf T_{\sigma}(f)$ via first-order uncertainty propagation from $\mathbf C_{\psi}$. Specifically, define
\begin{equation}
\mathbf h_{\psi}(f)\triangleq \mathrm{vec}\!\big(\widehat{\mathbf H}(f)\big),
\qquad
\mathbf J_{\psi}(f)\triangleq
\frac{\partial \mathbf h_{\psi}(f)}{\partial \boldsymbol{\psi}^{\mathsf T}}.
\end{equation}
Since $\widehat{\mathbf H}(f)\in\mathbb C^{L\times N_{\mathrm{TX}}}$, we partition
$\mathbf J_{\psi}(f)$ row-wise according to the receive-antenna index as
\begin{equation}
\mathbf J_{\psi}(f)
\triangleq
\begin{bmatrix}
\mathbf J_{\psi,1}(f)\\
\vdots\\
\mathbf J_{\psi,L}(f)
\end{bmatrix},
\qquad
\mathbf J_{\psi,\ell}(f)\in\mathbb C^{N_{\mathrm{TX}}\times d_{\psi}},
\end{equation}
where $\mathbf J_{\psi,\ell}(f)$ denotes the Jacobian block associated with the
$\ell$-th receive row of $\widehat{\mathbf H}(f)$, and
$d_{\psi}\triangleq \dim(\boldsymbol{\psi})$.
Then, $\mathbf T_{\sigma}(f)$ is approximated as
\begin{equation}
\mathbf T_{\sigma}(f)
\approx
\sum_{\ell=1}^{L}
\mathbf J_{\psi,\ell}(f)\,\mathbf C_{\psi}\,\mathbf J_{\psi,\ell}^{\mathsf H}(f),
\end{equation}
where $\mathbf C_{\psi}\in\mathbb C^{d_{\psi}\times d_{\psi}}$ and
$\mathbf T_{\sigma}(f)\in\mathbb C^{N_{\mathrm{TX}}\times N_{\mathrm{TX}}}$.

Under the subspace parameterization $\mathbf f=\mathbf W_{\mathrm{DT}}\boldsymbol{\eta}$,
both $\mu_{\Gamma}(\mathbf f)$ and $\sigma_{\Gamma}(\mathbf f)$ admit quadratic approximations with respect to $\boldsymbol{\eta}$.
Accordingly, the objective can be approximated as
\begin{equation}
\mu_{\Gamma}(\mathbf W_{\mathrm{DT}}\boldsymbol{\eta})
-\zeta_{\epsilon}\,\sigma_{\Gamma}(\mathbf W_{\mathrm{DT}}\boldsymbol{\eta})
\approx
\boldsymbol{\eta}^{\mathsf H}\mathbf G_{\mathrm{rob}}\boldsymbol{\eta},
\label{eq:Grob}
\end{equation}
with $\mathbf G_{\mathrm{rob}}\triangleq \mathbf G_{\mu}-\zeta_{\epsilon}\mathbf G_{\sigma}$, where $\mathbf G_{\mu}$ and $\mathbf G_{\sigma}$ denote the effective
subspace matrices induced by $\mathbf T_{\mu}$ and $\mathbf T_{\sigma}$, respectively.

Consequently, the second step of the closed-loop beam-control design reduces to the following robust \ac{BF} problem:
\begin{equation}
\max_{\boldsymbol{\eta}}\;
\boldsymbol{\eta}^{\mathsf H}\mathbf G_{\mathrm{rob}}\boldsymbol{\eta}
\quad \text{s.t.}\quad
\boldsymbol{\eta}^{\mathsf H}\mathbf E\,\boldsymbol{\eta} = 1 ,
\label{eq:bf_qp_robust}
\end{equation}
where
\begin{equation}
\mathbf E \triangleq \mathbf W_{\mathrm{DT}}^{\mathsf H}\mathbf W_{\mathrm{DT}}.
\end{equation}
This yields the generalized eigenvalue problem whose solution is given by the dominant generalized eigenvector. Specifically,
\begin{equation}
\boldsymbol{\eta}^\star
=
\operatorname{eig}_{\max}(\mathbf G_{\mathrm{rob}},\mathbf E),
\qquad
\mathbf f^\star =
\frac{\mathbf W_{\mathrm{DT}}\boldsymbol{\eta}^\star}
{\|\mathbf W_{\mathrm{DT}}\boldsymbol{\eta}^\star\|_2}.
\label{eq:bestBF_robust}
\end{equation}

From a control standpoint, the second step performs continuous beam refinement within the DT-informed subspace
spanned by the shortlisted codewords obtained in the preceding preselection step.
More broadly, the proposed closed-loop beam-control framework can be interpreted as a resource-reallocation mechanism:
by leveraging \ac{DT} priors to restrict the effective beam search space, the required pilot-signaling overhead is significantly reduced, while the complexity of \ac{DT} reconstruction can be handled offline.
In this sense, the framework trades offline computational resources for over-the-air training efficiency.

\section{Numerical Simulation Results}
\begin{figure}[t]
  \centering
  \includegraphics[width=0.6\columnwidth]{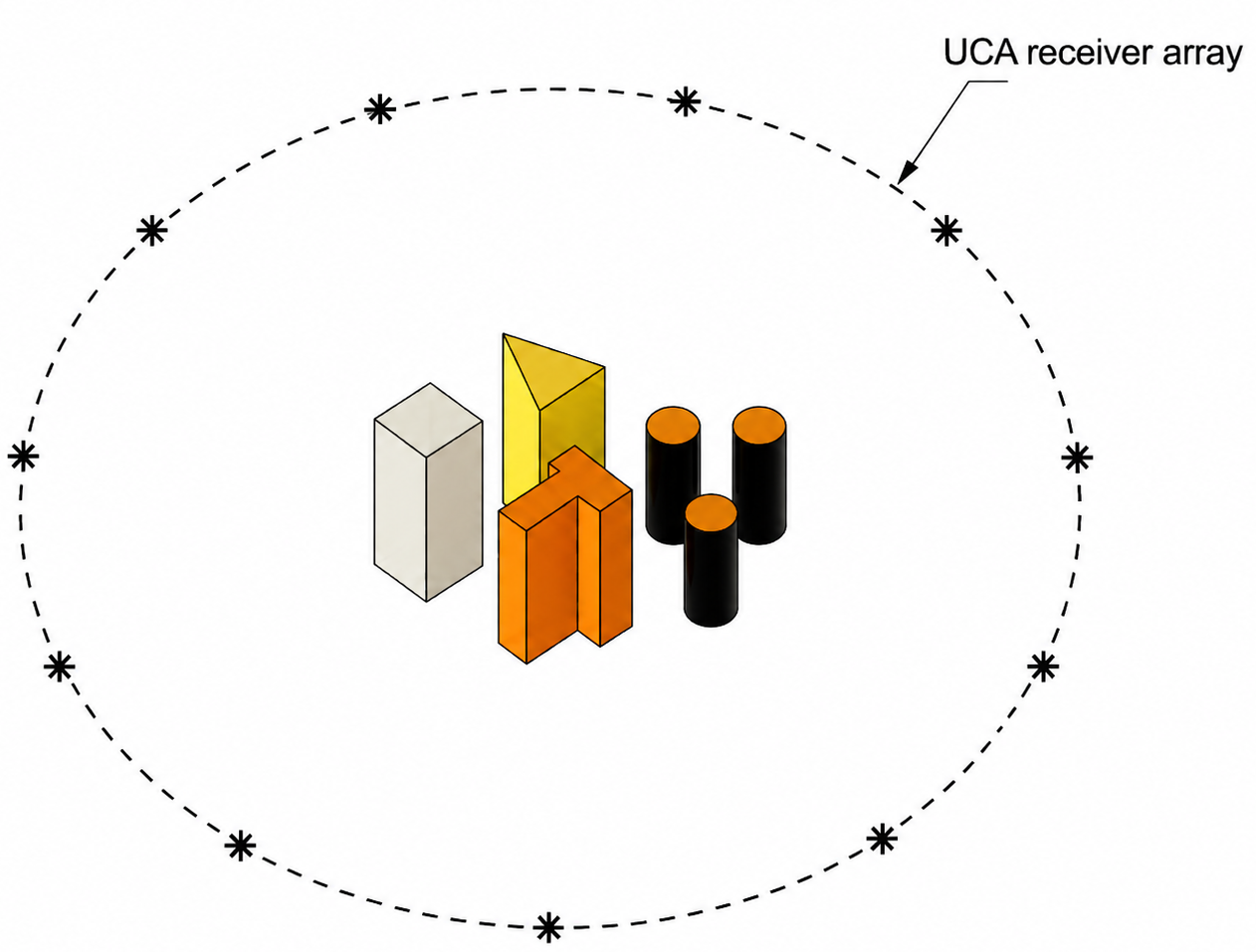}
  \caption{Illustration of the scatterer region.}
  \label{fig:cond}
\end{figure}

\subsection{Basic System Settings}
This section evaluates \ac{DT} construction under the \ac{GaBP} framework, including scatterer reconstruction, array calibration, and \ac{DT}-enabled closed-loop communication optimization.

As shown in Fig.~2, the sensing region is $1.56\,\mathrm{m}\times 1.56\,\mathrm{m}$ and contains an equilateral triangle, a square, a T-shaped object, and a cluster of three small circular scatterers. Their characteristic dimensions are $0.45\,\mathrm{m}$, $0.3\,\mathrm{m}$, $0.5\,\mathrm{m}$, and $0.1\,\mathrm{m}$, respectively.

To evaluate the ability of \ac{GaBP} to handle heterogeneous materials, three distinct material-parameter values are assigned to these scatterers. Each scatterer is assumed spatially homogeneous. Under the simplified TM$_z$ setting, all objects are uniform along the $z$-direction with height $6\,\mathrm{m}$.

The true antenna positions are generated by perturbing a nominal array geometry. The nominal RX array, also used as the inversion prior, is a \ac{UCA} with $16$ elements on a circle of radius $5\,\mathrm{m}$. The TX array is a \ac{ULA} with $18$ elements and half-wavelength spacing at $28\,\mathrm{GHz}$, located on the line $x=10$ and centered at $(10,2)$.

To emulate array-position errors, independent truncated Gaussian perturbations in polar coordinates are applied to each element in both radius and angle, with standard deviations set to $5\%$ of the nominal radius and $20^\circ$, respectively.

All electromagnetic field computations in the simulations are performed using the finite-difference time-domain (FDTD) method. Specifically, the sensing region $\mathcal{D}$ is discretized into a $52 \times 52$ grid. The OFDM carrier center frequency is set to $f_c = 28\,\mathrm{GHz}$, with $K = 36$ subcarriers and subcarrier spacing $\Delta f = 500\,\mathrm{kHz}$.

\subsection{Scatterer Material Reconstruction}
\begin{figure*}[t]
    \centering

    \subfloat[]{
        \includegraphics[width=0.22\linewidth]{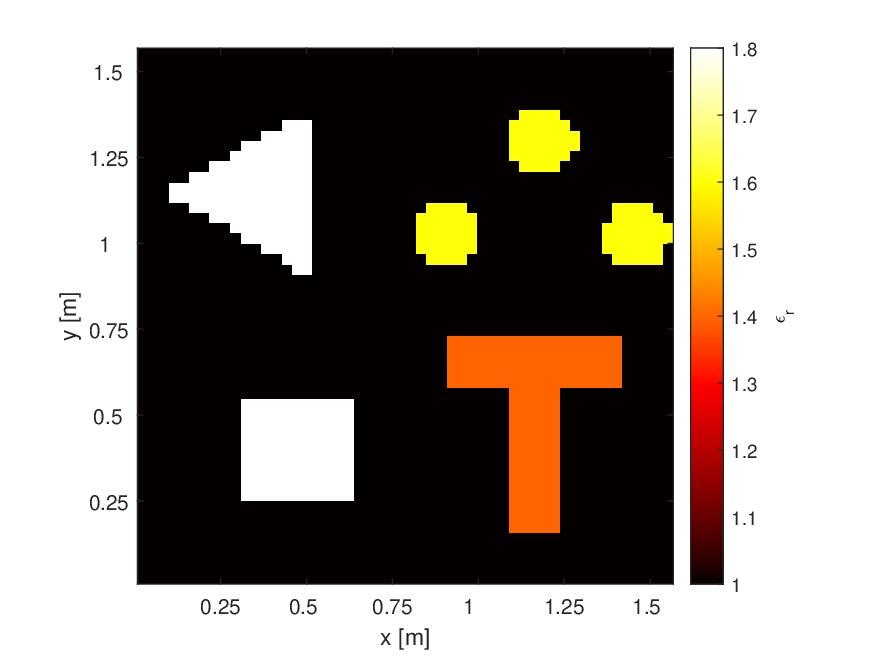}
    }
    \subfloat[]{
        \includegraphics[width=0.22\linewidth]{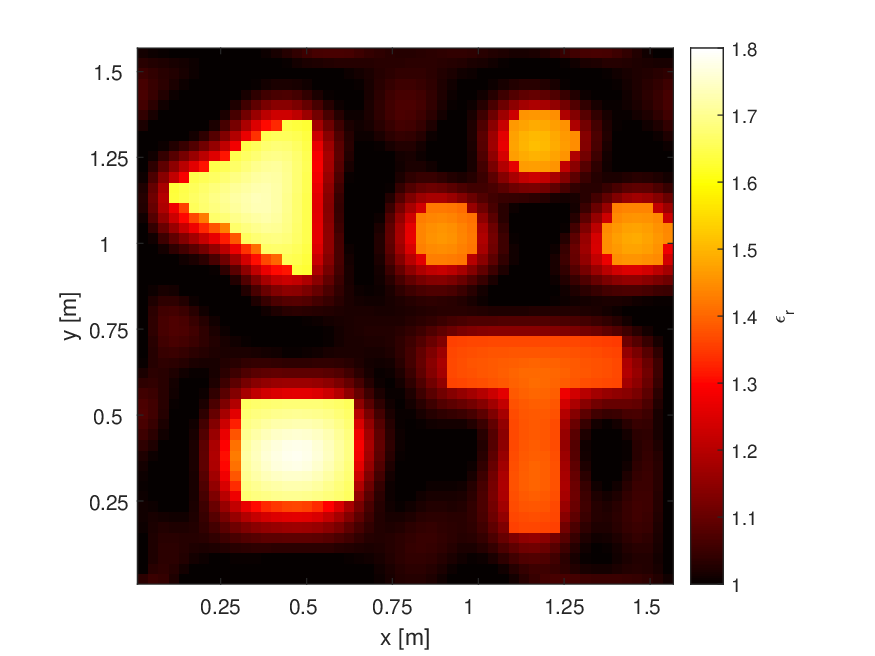}
    }
    \subfloat[]{
        \includegraphics[width=0.22\linewidth]{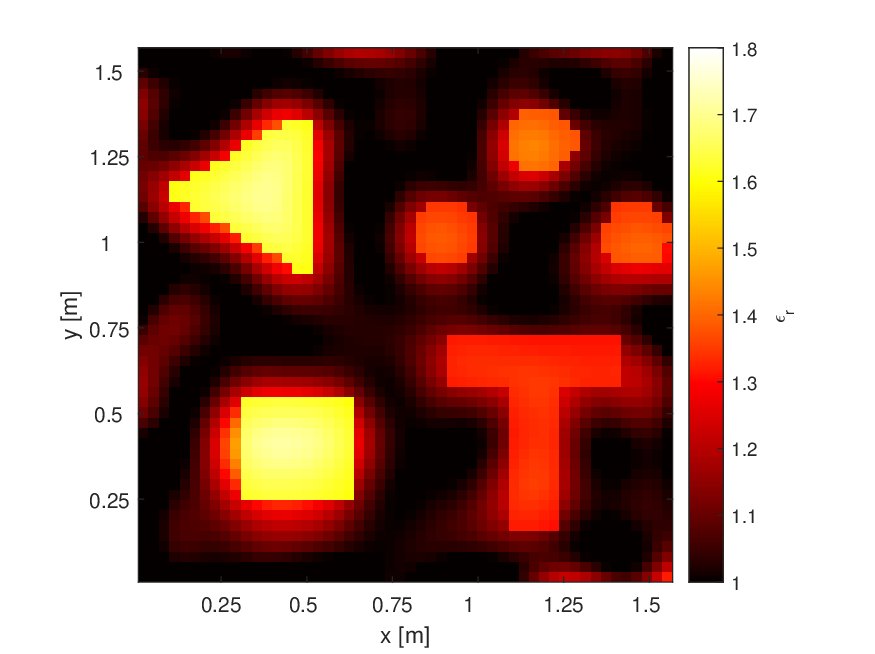}
    }\\

    \subfloat[]{
        \includegraphics[width=0.22\linewidth]{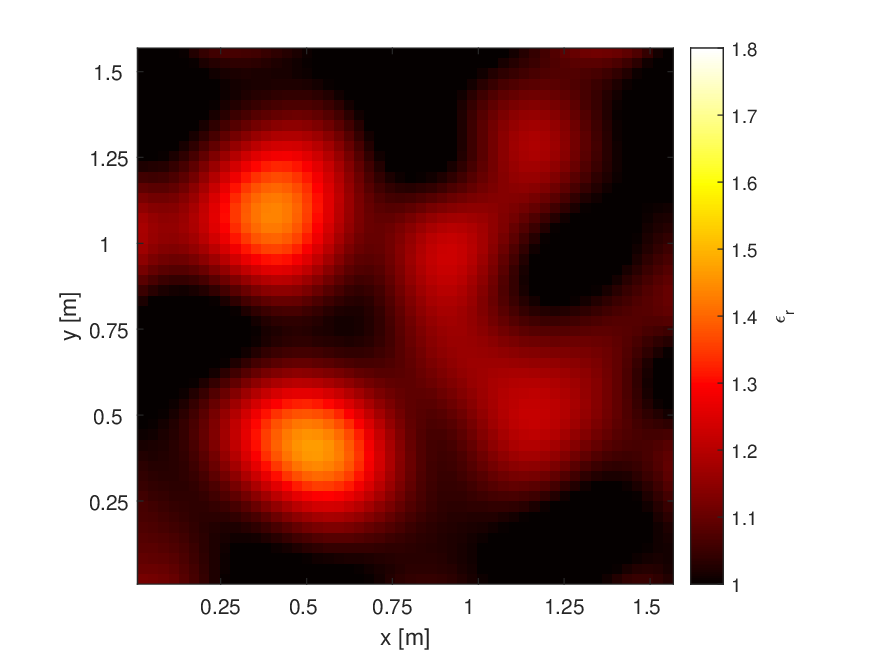}
    }
    \subfloat[]{
        \includegraphics[width=0.22\linewidth]{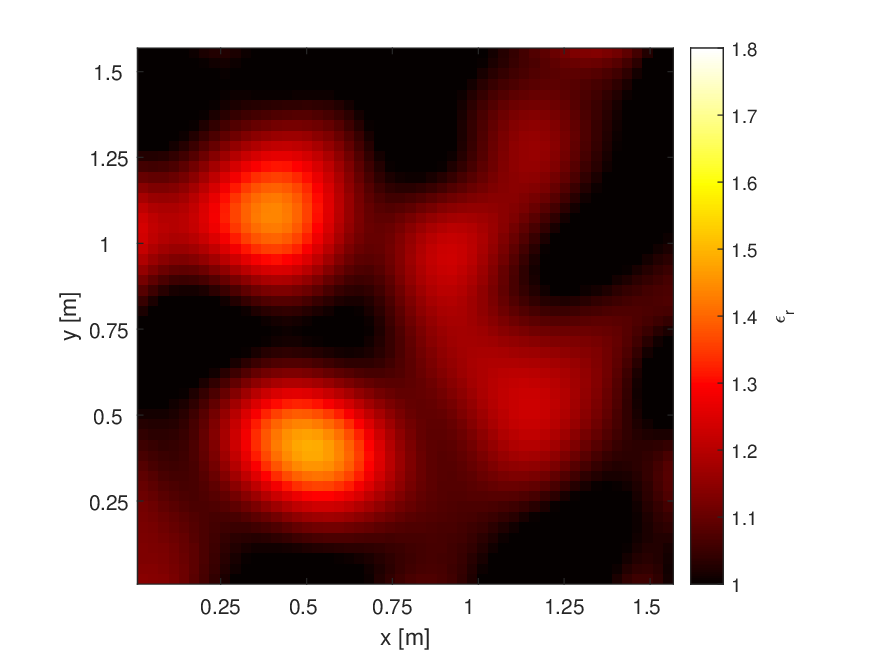}
    }
    \subfloat[]{
        \includegraphics[width=0.22\linewidth]{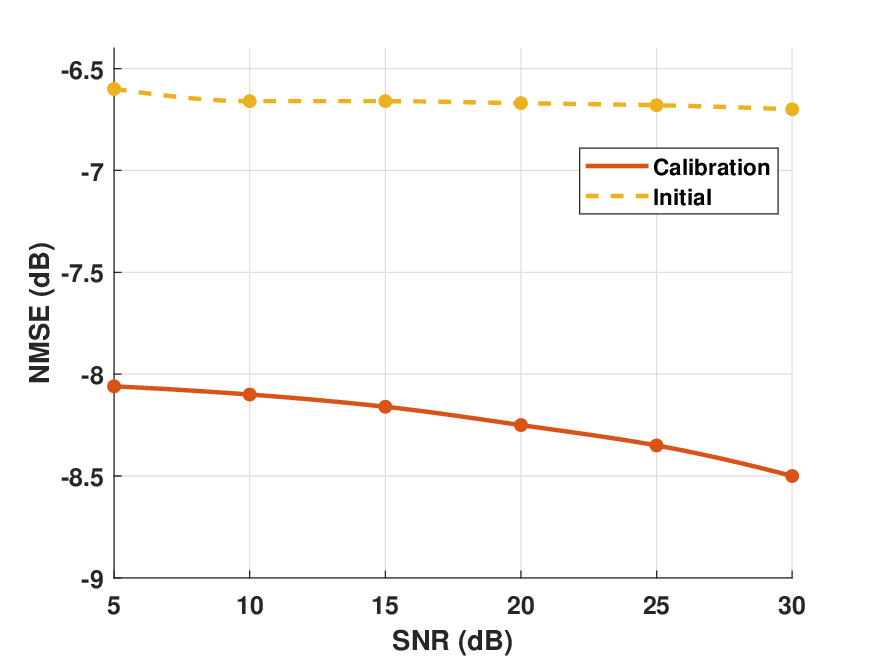}
    }
    \caption{Material reconstruction results under different sensing noise levels.
(a) ground-truth; (b)--(c) material reconstruction results with array calibration under sensing noise levels of $30~\mathrm{dB}$ and $5~\mathrm{dB}$, respectively; 
(d)--(e) material reconstruction results without array calibration under sensing noise levels of $30~\mathrm{dB}$ and $5~\mathrm{dB}$, respectively; (f) \ac{NMSE} performance.}
    \label{fig:reconstruction}
\end{figure*}

We evaluate material reconstruction under two sensing noise levels, $30~\mathrm{dB}$ and $5~\mathrm{dB}$, which characterize the noise intensity in the electromagnetic sensing process and are distinct from the communication-link \ac{SNR} defined later.

Fig.~\ref{fig:reconstruction}(a) shows the ground-truth 2-D material map: the triangle and square have $\varepsilon_r=1.8$, while the T-shape and the three-disk cluster have real-valued contrasts of $1.4$ and $1.6$, respectively.
Figs.~\ref{fig:reconstruction}(b) and (c) show the reconstructed maps obtained by the proposed \ac{GaBP} with array calibration at sensing noise levels of $30~\mathrm{dB}$ and $5~\mathrm{dB}$, respectively.

Under both weak-noise ($30~\mathrm{dB}$) and strong-noise ($5~\mathrm{dB}$) conditions, the proposed method accurately recovers the scatterer geometries while maintaining relatively small errors in the reconstructed material parameters. Note that antenna array calibration and material reconstruction are performed jointly in an alternating manner within the proposed \ac{GaBP} framework, and the results shown here correspond to the converged material estimates after the joint calibration--reconstruction process.

In contrast, as shown in Fig.~\ref{fig:reconstruction}(d) and (e), an unresolved antenna array mismatch causes the $\boldsymbol\chi$-step in the \ac{GaBP} algorithm to fail completely. In this case, the algorithm cannot recover the geometry or contrast distribution of any individual scatterer or scatterer cluster, and only spurious artifacts appear near the approximate scatterer locations. This behavior is caused by the use of an erroneous Green's function, which significantly worsens the ill-conditioning of the overall forward operator $\mathbf{A}$ and makes the solution of the associated linear system highly unstable. As a result, material reconstruction becomes infeasible under both weak- and strong-noise conditions.

Finally, Fig.~\ref{fig:reconstruction}(f) shows the NMSE performance of the $\boldsymbol\chi$-step in the \ac{GaBP} algorithm as a function of the sensing noise level for cases with and without antenna array calibration. The NMSE of the reconstructed $\boldsymbol{\chi}$ is defined as
\begin{equation}
\label{NMSE}
\mathrm{NMSE}_{\chi} \;=\; 10 \log_{10} \left(
\frac{\|\hat{\boldsymbol{\chi}} - \boldsymbol{\chi}\|_2^2}{\|\boldsymbol{\chi}\|_2^2}
\right).
\end{equation}

When antenna array calibration is incorporated, the $\boldsymbol\chi$-step of the \ac{GaBP} algorithm consistently achieves a low reconstruction error across different sensing noise levels, with the NMSE remaining around $-8~\mathrm{dB}$. As the sensing noise increases, the NMSE degrades by less than $0.5~\mathrm{dB}$, indicating strong noise robustness of the calibrated $\boldsymbol\chi$-step.

In contrast, in the presence of an unresolved antenna array mismatch, material reconstruction becomes ineffective because of the severe mismatch in the forward operator. Consequently, the NMSE remains nearly constant across different noise levels, further highlighting the necessity of antenna array calibration in the proposed \ac{GaBP} framework.

\begin{figure}[t]
  \centering
  \includegraphics[width=0.84\columnwidth]{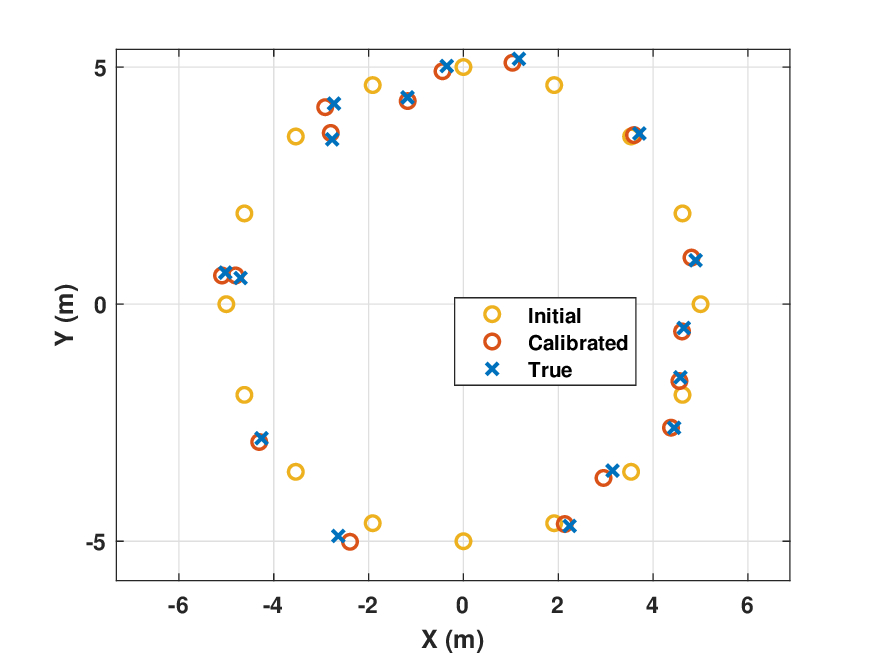}
  \caption{Calibrated array geometry.}
  \label{fig:arrayCal}
\end{figure}

\subsection{Array Calibration}

Following the material reconstruction results obtained in the $\boldsymbol\chi$-step, we next examine the performance of the proposed \texorpdfstring{$\boldsymbol{\theta}$-step}{theta-step}, which focuses on array calibration. In this and the subsequent subsection, all channel parameters are generated using the commercial ray-tracing simulator \textbf{Wireless InSite}, ensuring physically consistent propagation characteristics.

Fig.~\ref{fig:arrayCal} shows the antenna array calibration results obtained in the \texorpdfstring{$\boldsymbol{\theta}$-step}{theta-step} under a sensing noise level of $5~\mathrm{dB}$. The true antenna element positions are marked by blue crosses, the calibrated positions by red circles, and the prior (nominal) positions by orange circles. The estimated element locations closely match the true array geometry. After calibration, the NMSE of the antenna element positions relative to the true array geometry is reduced by $5.29~\mathrm{dB}$ in the radial domain and by $11.76~\mathrm{dB}$ in the angular domain, compared with the initial configuration.

The NMSE of the antenna array geometry is computed from the element-wise geometric deviations from the true array. Specifically, the NMSE in the radial and angular domains is defined as 
$\mathrm{NMSE}_r = 10 \log_{10} \!\left(
\frac{\sum_{\ell=1}^{L} (\hat r_\ell - r_\ell)^2}
{\sum_{\ell=1}^{L} r_\ell^2}
\right)$ 
and 
$\mathrm{NMSE}_\phi = 10 \log_{10} \!\left(
\frac{\sum_{\ell=1}^{L} (\hat \phi_\ell - \phi_\ell)^2}
{\sum_{\ell=1}^{L} \phi_\ell^2}
\right)$,
where $r_\ell$ and $\phi_\ell$ denote the true radial distance and angular position of the $\ell$-th antenna element, respectively, while $\hat r_\ell$ and $\hat \phi_\ell$ are the corresponding calibrated estimates.

To further assess the accuracy of the calibrated antenna array at the channel level, the \ac{AoA} and the \ac{PDP} are adopted as performance metrics.

\begin{figure}[t]
    \centering
    \subfloat[Calibrated \ac{AoA} spectrum.]{
        \includegraphics[width=0.47\linewidth]{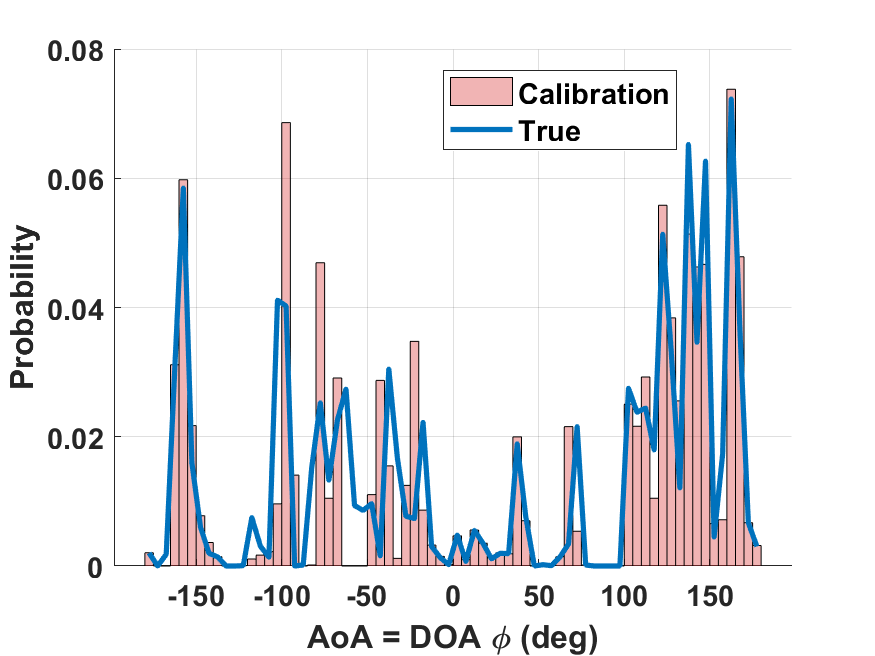}
    }
    \hfill
    \subfloat[Uncalibrated \ac{AoA} spectrum.]{
        \includegraphics[width=0.47\linewidth]{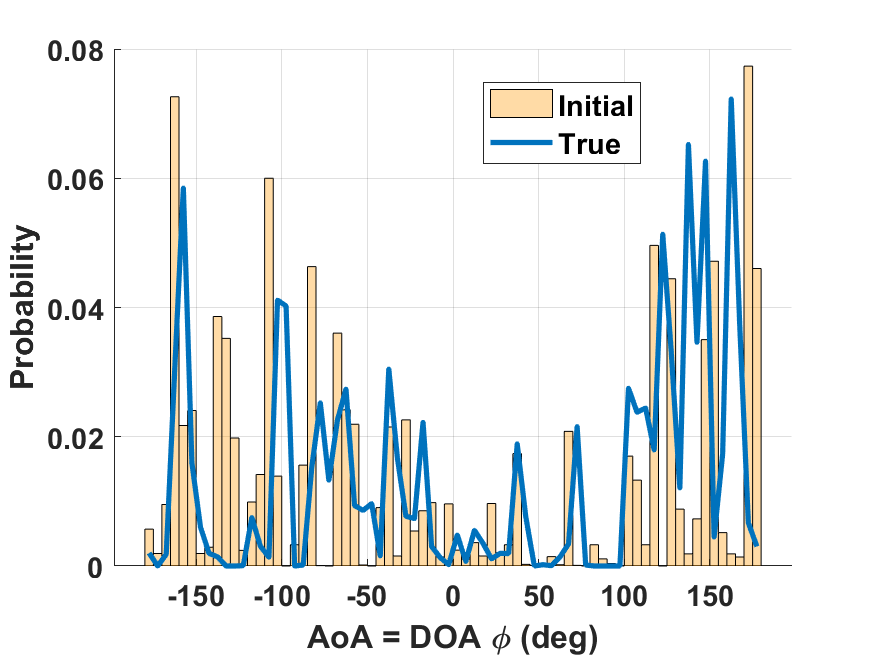}
    }\\
    \subfloat[Calibrated \ac{AoA} error \ac{CDF}.]{
        \includegraphics[width=0.47\linewidth]{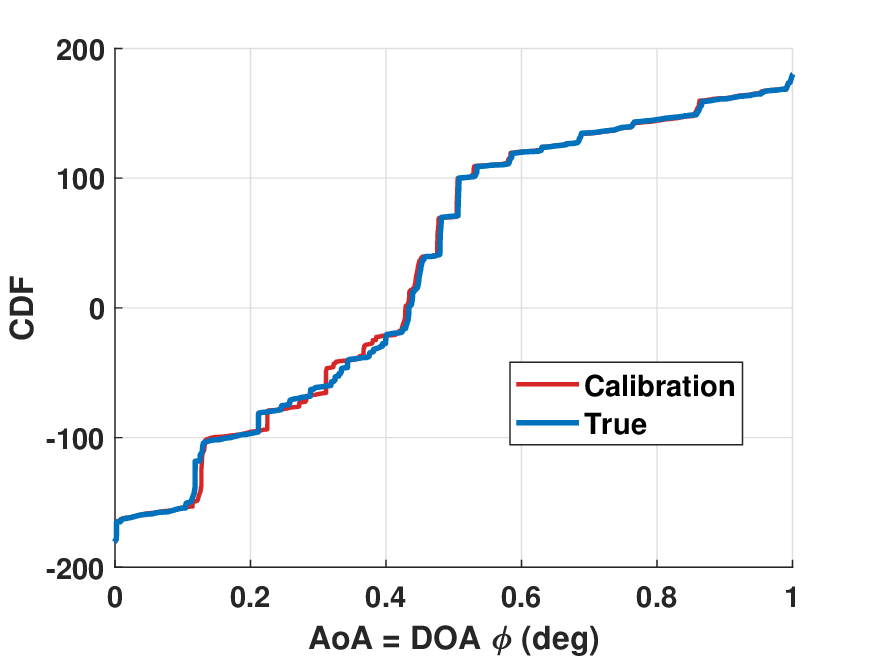}
    }
    \hfill
    \subfloat[Uncalibrated \ac{AoA} error \ac{CDF}.]{
        \includegraphics[width=0.47\linewidth]{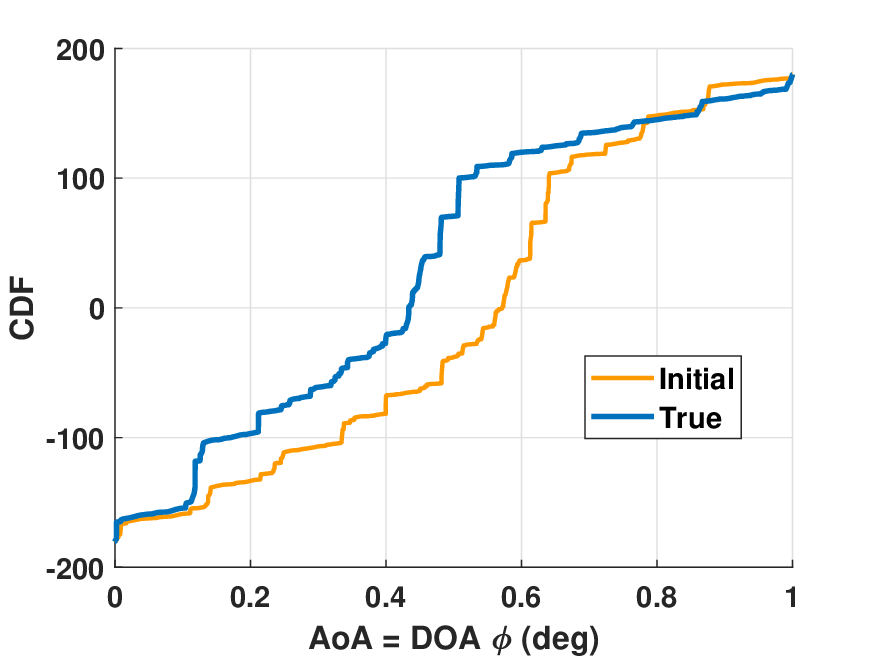}
    }
    \caption{\ac{AoA} estimation results and corresponding error \acp{CDF} with and without array calibration.}
    \label{fig:aoa_all}
\end{figure}

As shown in Fig.~\ref{fig:aoa_all}(a), the \ac{AoA} spectrum after calibration agrees much better with the reference distribution. Near the dominant component around $-100^\circ$, the calibrated spectrum matches both the reference peak location and its local shape, indicating accurate recovery of the main arriving direction. By contrast, the uncalibrated result in Fig.~\ref{fig:aoa_all}(b) is clearly distorted, with inconsistent peak prominence and spurious nearby components caused by array mismatch. In the $100^\circ$--$180^\circ$ sector, where several strong components cluster and the spectrum becomes more structured, the calibrated curve preserves the reference envelope and peak concentration pattern, whereas the uncalibrated spectrum exhibits a clear shape mismatch, including exaggerated peaks and peak spreading. Overall, calibration substantially improves channel-level \ac{AoA} fidelity, especially near $-100^\circ$ and over $100^\circ$--$180^\circ$.

These gains are further supported by the \ac{AoA} error \acp{CDF} in Fig.~\ref{fig:aoa_all}(c) and Fig.~\ref{fig:aoa_all}(d). With calibration, the \ac{CDF} nearly overlaps with the reference over the full range, particularly in the low-to-mid error region (roughly below $0.5^\circ$), indicating that most estimates have small angular deviations. Without calibration, the \ac{CDF} shifts toward larger errors, with a clear gap in the mid-quantiles (about $0.4^\circ$--$0.7^\circ$) and a heavier tail, indicating more frequent large-angle outliers. Overall, array calibration suppresses large-error events and significantly improves \ac{AoA} reliability.

\begin{figure}[t]
    \centering
    \subfloat[Calibrated \ac{PDP}.]{
        \includegraphics[width=0.47\linewidth]{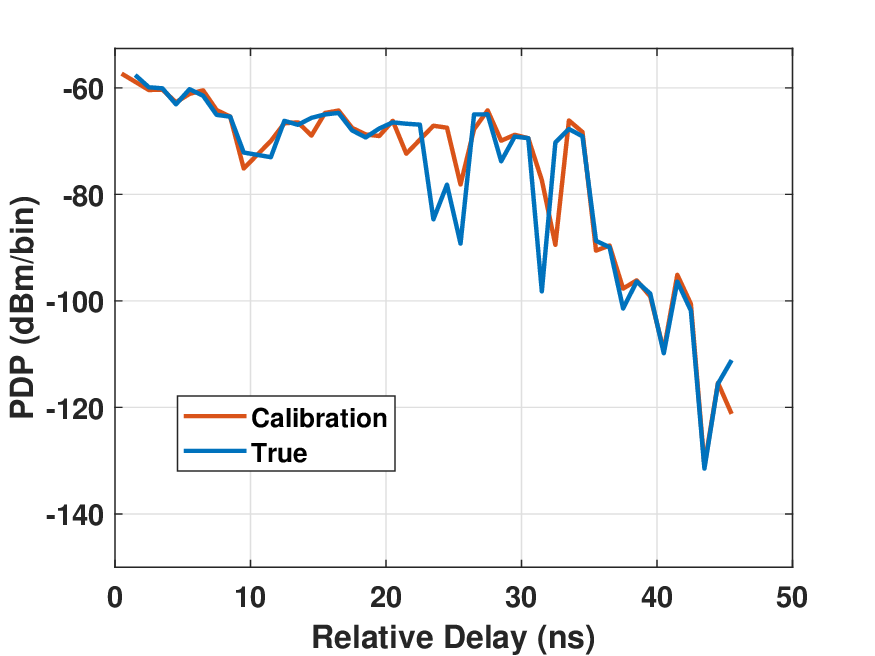}
    }
    \hfill
    \subfloat[Uncalibrated \ac{PDP}.]{
        \includegraphics[width=0.47\linewidth]{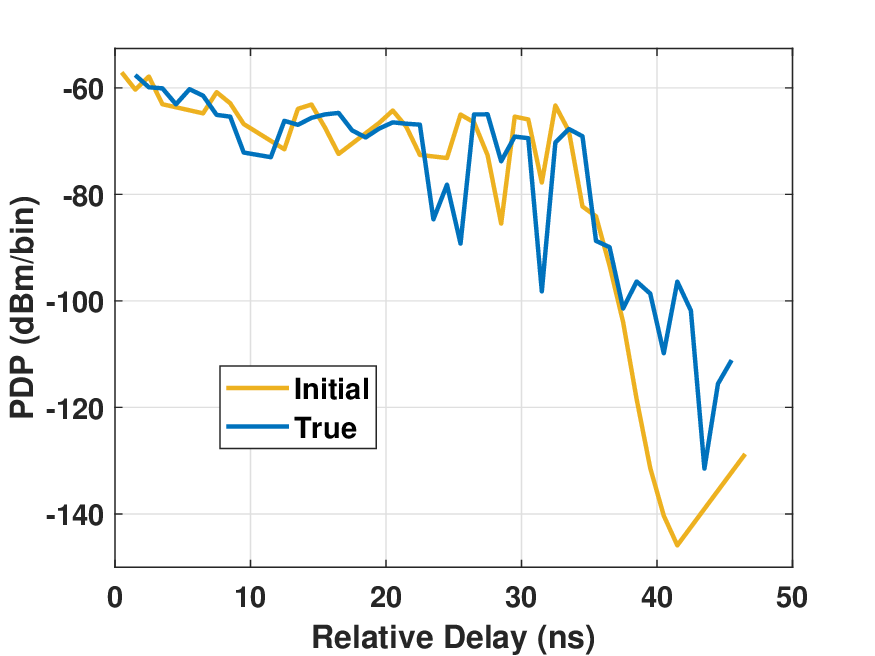}
    }\\
    \subfloat[Calibrated \ac{PDP} error \ac{CDF}.]{
        \includegraphics[width=0.47\linewidth]{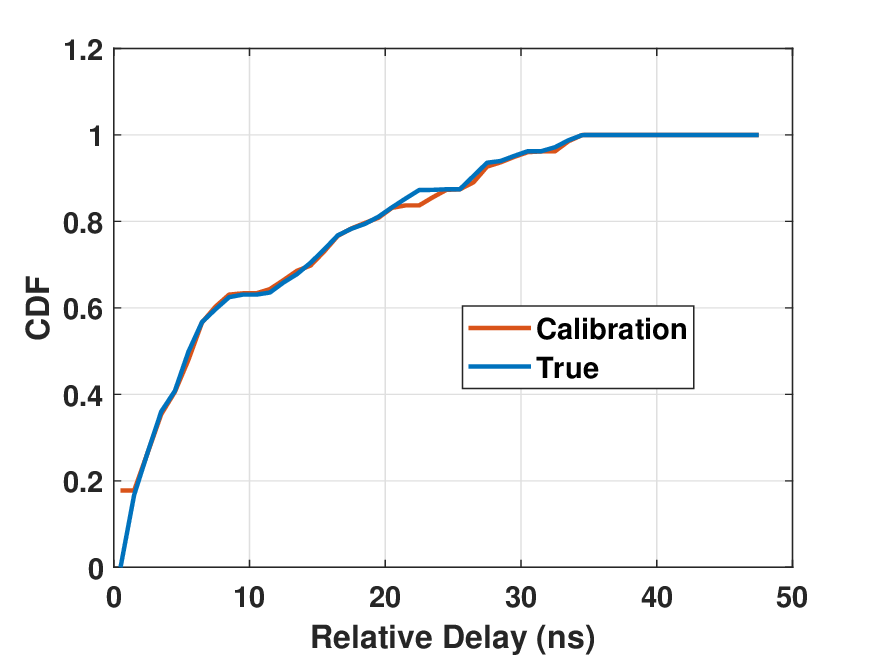}
    }
    \hfill
    \subfloat[Uncalibrated \ac{PDP} error \ac{CDF}.]{
        \includegraphics[width=0.47\linewidth]{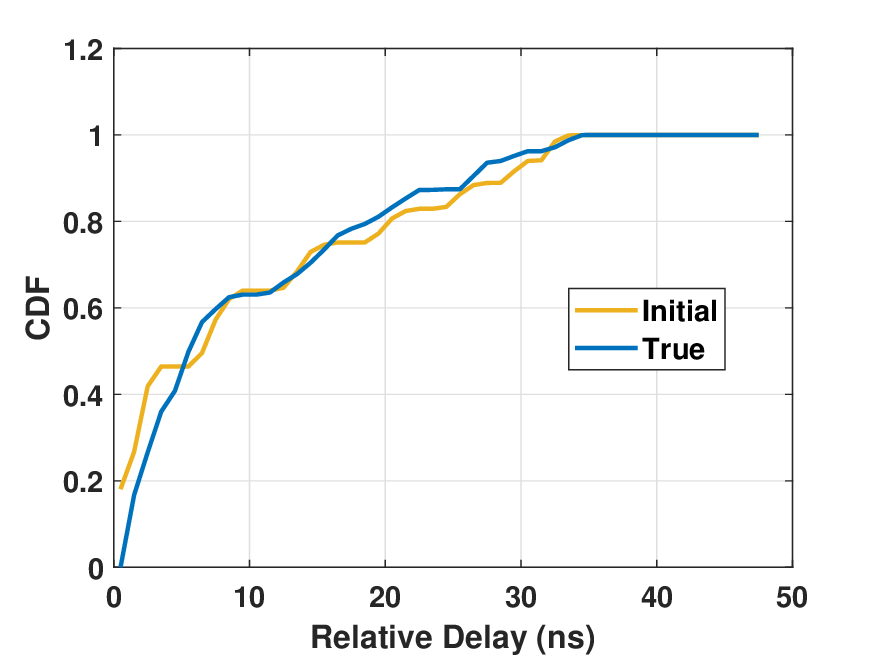}
    }
    \caption{\ac{PDP} estimation results and corresponding error \acp{CDF} with and without array calibration.}
    \label{fig:pdp_all}
\end{figure}

As shown in Fig.~\ref{fig:pdp_all}(a), the calibrated \ac{PDP} closely matches the ground-truth curve over most delay intervals. In the main multipath region ($0$--$35~\mathrm{ns}$), the two trajectories largely overlap, with typical deviations of about $2$--$3~\mathrm{dB}$ across most delay bins. At larger delays (around $40$--$47~\mathrm{ns}$), discrepancies remain but are generally limited to about $10~\mathrm{dB}$, with the largest local deviations reaching roughly $10$--$15~\mathrm{dB}$.

Without array calibration, as shown in Fig.~\ref{fig:pdp_all}(b), the \ac{PDP} becomes strongly distorted in the delay tail. Around $40$--$45~\mathrm{ns}$, the uncalibrated \ac{PDP} exhibits an artificial deep fade, dropping to about $-145~\mathrm{dB}$ while the ground truth stays near $-125$ to $-130~\mathrm{dB}$, i.e., a local error of roughly $15$--$20~\mathrm{dB}$. This suggests that array mismatch can severely corrupt multipath-energy estimation, especially at large delays.

The error \acp{CDF} in Fig.~\ref{fig:pdp_all}(c) and Fig.~\ref{fig:pdp_all}(d) further corroborate this trend. With calibration, the \acp{CDF} nearly coincide, and the horizontal gaps at the $0.8$ and $0.9$ quantiles are typically within $1$--$2~\mathrm{ns}$. Without calibration, the \ac{CDF} shifts toward larger errors; for instance, near the $0.8$ quantile the gap increases to about $3$--$5~\mathrm{ns}$, indicating a heavier error tail in the uncalibrated case.

\subsection{DT-Guided Codebook Preselection}

\begin{figure}[t]
  \centering
  \includegraphics[width=.68\columnwidth]{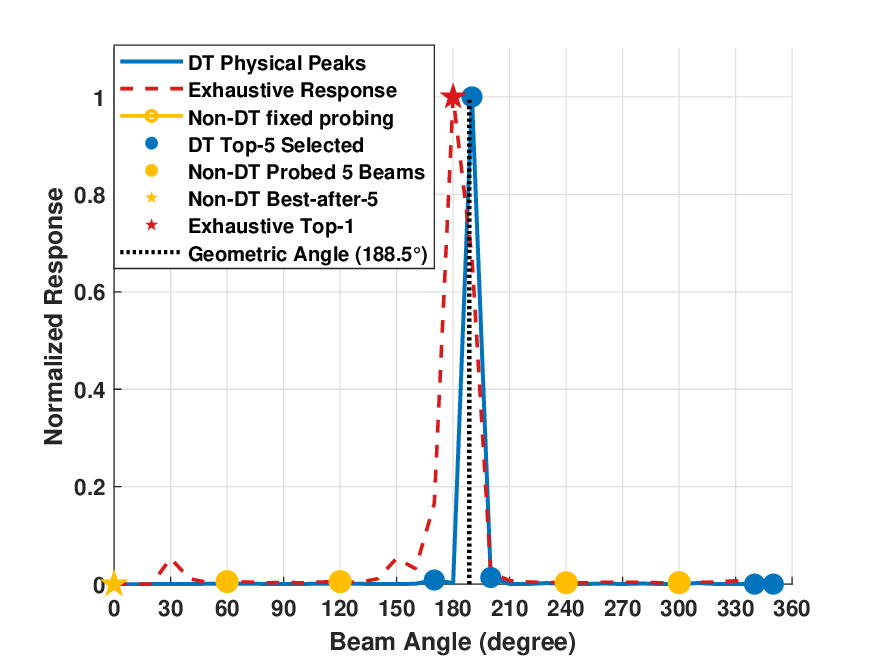}
  \caption{Comparison of codeword selection results.}
  \label{fig:beamselect}
\end{figure}

After the joint iterations of the $\boldsymbol\chi$-step and the \texorpdfstring{$\boldsymbol{\theta}$-step}{theta-step} have converged and reliable environment and array estimates are obtained, the system proceeds to the DT-guided codebook preselection stage. This subsection evaluates the effectiveness of the proposed DT-assisted codebook preselection strategy and compares it with the conventional exhaustive-search approach.

In the simulations, a uniform angular-domain codebook is adopted, where the beam directions are uniformly sampled from $0^\circ$ with a step size of $10^\circ$, resulting in a total of $36$ codewords. For the exhaustive-search method, pilot signals are sequentially transmitted along all codeword directions in the codebook, and the codeword corresponding to the maximum received power is selected. In contrast, the \ac{DT}-based method avoids full angular scanning by exploiting the prior geometric and environmental information provided by the constructed \ac{DT}. Specifically, according to \eqref{eq:beamselect}, a subset of the most relevant codewords is directly selected from the codebook. In the following experiments, the number of selected candidate codewords is set to $B=5$.

Fig.~\ref{fig:beamselect} compares the codeword selection outcomes of the \ac{DT}-assisted strategy and the conventional exhaustive-search benchmark.
For the \ac{DT}-based method, \eqref{eq:beamselect} yields a candidate set with $B=5$ codewords that captures the angular support inferred from the constructed \ac{DT}.
In the considered scenario, where the scattering energy is highly concentrated in angle, only the codeword pointing at $190^\circ$ within the candidate set corresponds to the dominant propagation direction, while the remaining candidates mainly serve to provide robustness against angular quantization and modeling uncertainties.
By contrast, under the same $B=5$ air-interface budget, a \emph{No-DT} baseline follows a fixed probing rule (three uniformly spaced beams and two midpoint refinements based on the summed feedback energy), which relies on coarse statistical priors and thus may miss the dominant direction.
The exhaustive-search benchmark selects a codeword pointing at $180^\circ$.
Notably, the two selected directions differ by only one codeword step under the adopted $10^\circ$ angular resolution and therefore correspond to very similar beam directions with comparable link-level performance.
Moreover, the geometric angle of the scatterer-region centroid relative to the TX array center is approximately $188.5^\circ$, indicating that the codeword selected by the \ac{DT}-based method is more consistent with the underlying scattering geometry.
Overall, this result demonstrates that reliable codebook preselection can be achieved by exploiting \ac{DT}-derived priors while operating on a significantly reduced candidate set.

\subsection{Closed-Loop Beam Management}

\begin{figure}[t]
  \centering
  \includegraphics[width=.68\columnwidth]{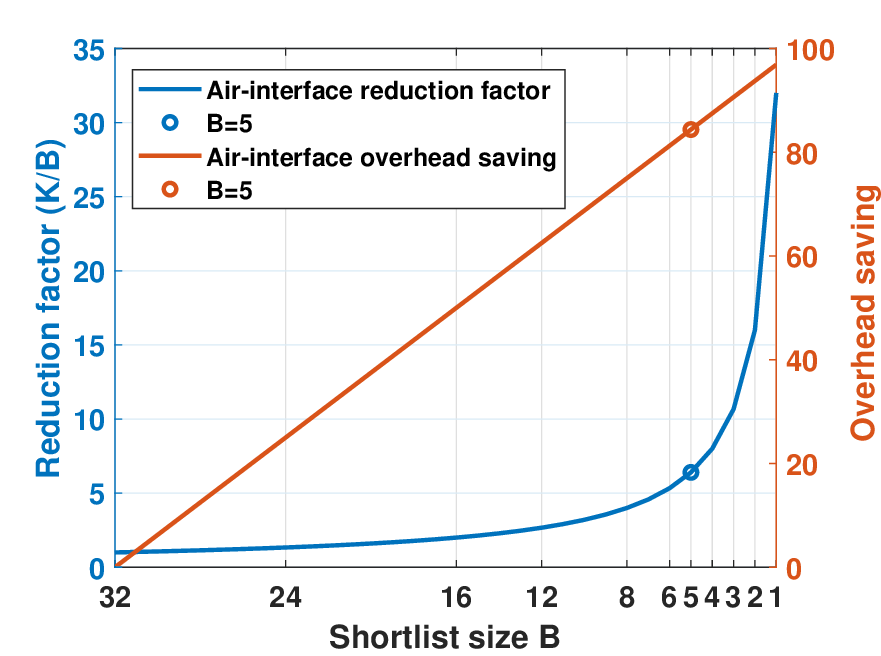}
  \caption{Impact of shortlist size $B$ on air-interface overhead.}
  \label{fig:Btrend}
\end{figure}

\begin{figure}[t]
  \centering
  \includegraphics[width=.68\columnwidth]{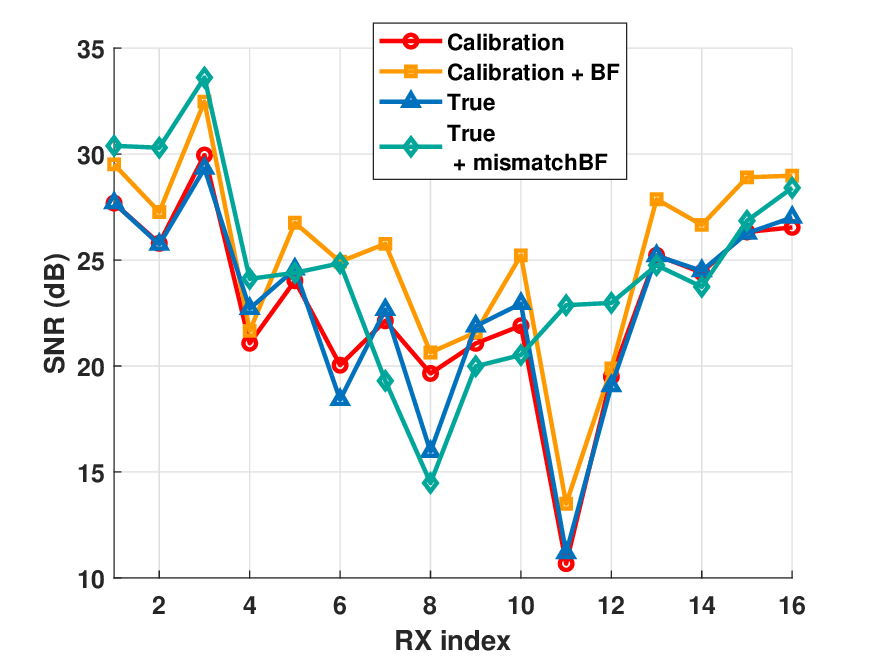}
  \caption{\ac{SNR} trends across individual RX antennas under different \ac{BF} configurations.}
  \label{fig:SNR}
\end{figure}

\begin{figure}[t]
  \centering
  \includegraphics[width=.68\columnwidth]{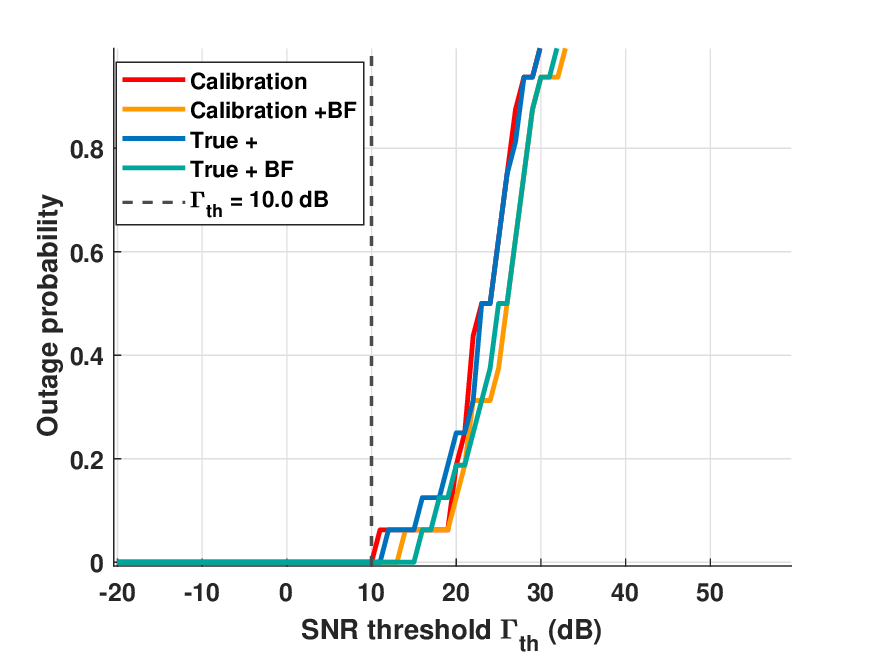}
  \caption{Outage probability versus SNR threshold under different \ac{BF} configurations.}
  \label{fig:outage}
\end{figure}

After obtaining the DT-guided shortlisted codewords in the preceding subsection, we perform transmit beam optimization within this reduced candidate set, thereby completing the subsequent closed-loop beam-control procedure.

The gain of the proposed DT-guided strategy lies in an explicit \emph{resource tradeoff}: offline computation for DT construction is used to reduce the over-the-air beam-sweeping burden. Under our experimental setting, DT construction takes approximately $10$~min, and the corresponding air-interface overhead reduction is depicted in Fig.~\ref{fig:Btrend}.

We quantify the beam-training cost by the \emph{number of swept beams per beam-management cycle}, i.e., the number of codewords that must be probed on the air interface to identify a suitable transmit beam. As shown in Fig.~\ref{fig:Btrend}, decreasing the shortlist size $B$ leads to larger savings: the sweeping overhead scales with $B$, whereas the reduction factor follows $U/B$, where $U=36$ is the codebook size. In our implementation, the baseline probes all $36$ codewords, while the DT-guided scheme probes only $B=5$ shortlisted codewords for on-air confirmation, yielding a $7.2$ times reduction factor and an $86.11\%$ overhead saving. These results indicate that DT-guided inference can markedly reduce over-the-air sweeping while maintaining near-oracle \ac{BF} performance.

Furthermore, to demonstrate the gain of DT-guided \ac{BF}, we evaluate the received \ac{SNR} and outage probability.
Fig.~\ref{fig:SNR} shows the per-antenna received \ac{SNR} at the RX array for the calibrated and true arrays, with and without \ac{BF}.
Except for RX antenna index $12$, \ac{BF} improves the \ac{SNR} by about $1$--$2.5~\mathrm{dB}$ across most antennas.
The \ac{SNR} curves of the calibrated array closely match those of the true array, consistent with the high calibration accuracy reported earlier.
For comparison, the green curve represents \ac{BF} based on the original uncalibrated array geometry.
Although it improves the \ac{SNR} at some antennas, it also causes clear degradation at others.
For example, at RX antenna index $8$, the \ac{SNR} decreases by $1.48~\mathrm{dB}$ compared with the no-\ac{BF} case.
This confirms the importance of accurate array geometry for spatially consistent \ac{BF} performance.

Fig.~\ref{fig:outage} reports the outage probability as a function of the \ac{SNR} threshold $\Gamma_{\mathrm{th}}$, where the dashed line highlights $\Gamma_{\mathrm{th}}=10~\mathrm{dB}$. It can be observed that the outage probability at $\Gamma_{\mathrm{th}}=10~\mathrm{dB}$ is essentially zero for all configurations, indicating that the link reliably exceeds this threshold in the considered setup. As $\Gamma_{\mathrm{th}}$ increases into the transition region (roughly $25$--$35~\mathrm{dB}$), applying \ac{BF} consistently shifts the outage curves to the right, i.e., a lower outage is achieved at the same threshold. For example, around $\Gamma_{\mathrm{th}} \approx 30~\mathrm{dB}$, \ac{BF} reduces the outage probability by roughly $0.05$--$0.15$ compared with the corresponding no-\ac{BF} cases for both the calibrated-array and true-array settings. Moreover, the outage curves of the calibrated-array cases closely track those of the true-array cases, again confirming that the calibrated array yields channel-level performance that is highly consistent with the true array geometry. By contrast, \ac{BF} based on the uncalibrated array geometry exhibits less consistent outage behavior because the reduced \ac{SNR} at certain antennas increases the probability of falling below higher \ac{SNR} thresholds.

\section{Conclusion}

This paper proposed an electromagnetic \ac{ISAC}-assisted digital twin framework for antenna array calibration and DT-guided \ac{BF}. By reconstructing a physically consistent digital twin that captures both the propagation environment and array geometry, the proposed approach mitigates array-manifold mismatch and supports more reliable model-based \ac{BF}. More importantly, the proposed closed-loop architecture provides a resource-reallocation mechanism: rather than relying mainly on exhaustive over-the-air training, it leverages offline electromagnetic reconstruction and digital-twin inference to narrow the beam-management space and support downstream communication optimization. In this sense, the method effectively trades offline computational resources for air-interface resources, reducing online training overhead while maintaining robust \ac{BF} performance. Overall, the results highlight the value of accurate DT reconstruction as a bridge from electromagnetic sensing to actionable communication control, thereby establishing a closed-loop link among environment perception, digital twin construction, and wireless system optimization.

\end{document}